\documentclass[
aps,
prx,
showpacs,
preprintnumbers,
twocolumn,
superscriptaddress,
longbibliography,
10pt,
]{revtex4-2}

\usepackage[T1]{fontenc}
\usepackage{newtxtext}
\usepackage{newtxmath}

\usepackage{enumitem}

\usepackage{physics}
\usepackage{amsmath}
\usepackage{mathtools}
\usepackage{dsfont}

\usepackage{appendix}

\usepackage{placeins}

\usepackage{color}
\usepackage[dvipsnames]{xcolor}

\definecolor{nqdcolor}{rgb}{0.5586, 0.0586, 0.4219}

\usepackage{hyperref}
\hypersetup{colorlinks,citecolor=nqdcolor,linkcolor=nqdcolor,urlcolor=nqdcolor}

\usepackage{makecell}
\usepackage{colortbl} 

\usepackage{graphicx}
\usepackage[all]{hypcap}

\usepackage{microtype}
\usepackage{tikz}

\newcommand{\tcav}{\tau_{\rm cav}}
\newcommand{\tosc}{\tau_{\rm osc}}
\newcommand{\tcorr}{\tau_{\rm corr}}
\newcommand{\tcorrens}{\tau_{\rm corr}^{\rm ens}}
\newcommand{\tth}{\tau_{\rm th}}
\newcommand{\Tbath}{T_{\rm bath}}
\newcommand{\omgc}{\omega_{\rm c}}
\newcommand{\nsol}{\rho_{\rm sol}}
\newcommand{\tp}{\tau_{\rm }}

\newcommand{\nocontentsline}[3]{}
\let\origcontentsline\addcontentsline
\newcommand\stoptoc{\let\addcontentsline\nocontentsline}
\newcommand\resumetoc{\let\addcontentsline\origcontentsline}

\makeatletter
\@ifundefined{ifmainmode}{\newif\ifmainmode\mainmodetrue}{}%
\@ifundefined{ifsuppmode}{\newif\ifsuppmode\suppmodetrue}{}%
\@ifundefined{ifcombinedmode}{\newif\ifcombinedmode\combinedmodetrue}{}%
\makeatother

\newcommand{\PKS}{Max Planck Institute for the Physics of Complex Systems, MPI-PKS, N\"othnitzerstr.~38, 01187 Dresden, Germany}
\newcommand{\SCNU}{Institute for Theoretical Physics, School of Physics, South China Normal University, Guangzhou 510006, China}
\newcommand{\SCNULAB}{Guangdong Provincial Key Laboratory of Quantum Engineering and Quantum Materials, Guangdong-Hong Kong Joint Laboratory of Quantum Matter, South China Normal University, Guangzhou 510006, China}

\begin{document}

\title{Discrete Time Liquids}

\author{Zuo Wang}
\email[Correspondence to: ]{zuowang@pks.mpg.de}
\affiliation{\PKS}\affiliation{\SCNU}\affiliation{\SCNULAB}
\author{Paul M.~Schindler}\affiliation{\PKS}
\author{Marin Bukov}\affiliation{\PKS}
\author{Marios H.~Michael}
\email[Correspondence to: ]{mariosm@pks.mpg.de}
\affiliation{\PKS}

\begin{abstract}
Discrete time crystals exhibit robust subharmonic oscillations, spontaneously breaking the discrete time-translation symmetry of a periodic drive. Since their emergence, the question of whether other types of temporal many-body order can exist has attracted considerable attention. Here we identify discrete time liquids -- a new form of spatiotemporal order characterized by short-range temporal coherence as a result of time-solitons. We study a two-dimensional spin system coupled to a finite-temperature cavity, whose fluctuations provide a noisy, approximately periodic drive with only finite temporal coherence. Remarkably, the spins nevertheless form a metastable state with robust period-doubled dynamics over an extended parameter regime. Unlike a time crystal, this state has only finite temporal coherence: individual trajectories stochastically switch between the two symmetry-related temporal configurations of the period-doubled response. We identify these switches as time-solitons: topological defects in temporal order that preserve local period doubling while destroying long-range temporal coherence, thereby producing the time liquid state. We propose a minimal model which reproduces the distribution of temporal correlations of the many-body system, suggesting a general description of the time liquid as a time crystal with a finite density of time-solitons. Because these solitons can occur at arbitrary times, ensemble averaging restores continuous time-translation symmetry, raising the intriguing possibility that discrete time liquids can exist in thermal equilibrium.
\end{abstract}

\maketitle
	

\section{Introduction}

Fluctuations are usually regarded as an enemy of order. In spatially ordered phases, they blur static patterns and can forbid long-range order in low dimensions~\cite{Mermin1966PRL,Hohenberg1967PR}; in dynamical systems, they dephase coherent motion and broaden sharp spectral features~\cite{Lax1967PR,Pikovsky2001Book}. Nevertheless, fluctuations can also play a constructive role. In order by disorder, fluctuations select ordered states that are not preferred at the mean-field level~\cite{Villain1980JPF,Henley1989PRL}. In stochastic and coherence resonances, fluctuations optimize the response of nonlinear dynamical systems~\cite{Benzi1981JPA,Gammaitoni1998RMP,Pikovsky1997PRL}. Both of these cases show that fluctuations can also create regularity rather than merely destroy it. This idea has recently expanded into the field of cavity-matter hybrids~\cite{bretscher26,Eckhardt25,Emil25,Keren26}. There, the goal is to generate new phases of matter by amplifying electromagnetic fluctuations around a two-dimensional heterostructure through nanopatterning of the electromagnetic environment. In this article, we find that fluctuations not only generate new spatially coherent phases of matter, but can also give rise to many-body temporal order, producing what we term \textit{discrete time liquids}.  

Over the past decade, quantum many-body systems have been shown to support robust phases of matter with temporal order known as time crystals. In a conventional discrete time crystal (DTC), an externally imposed periodic drive defines a discrete time-translation symmetry, which the many-body system spontaneously breaks by locking local observables to a robust subharmonic response~\cite{Zaletel2023RMP}. Such order has been established in many-body-localized, prethermal, classical Floquet and driven-dissipative systems~\cite{Sacha2018RPP,Else2016PRL,Khemani2016PRL,Yao2017PRL,Choi2017Nature,Else2017PRX,Kyprianidis2021Science,Yao2020NatPhys,Iemini2018PRL,Gong2018PRL,Gambetta2019PRL,Kessler2021PRL,Schumann2026PRL}. In most realizations, however, the temporal structure is supplied by an external drive that is, to a good approximation, perfectly periodic. A schematic time-crystal evolution for ferromagnetically interacting spins is generated over one period of unit length by
\begin{equation}
\hat U = e^{i J\sum_{\langle ij\rangle}\hat{S}^z_i\hat{S}^z_j} e^{i\pi\sum_i\hat{S}^x_i},
\label{eq:DTCequation}
\end{equation}
which combines ferromagnetic interactions with a global spin flip. The periodic drive therefore provides a coherent clock against which the subharmonic response is defined.

\begin{figure*}
\centering
\includegraphics{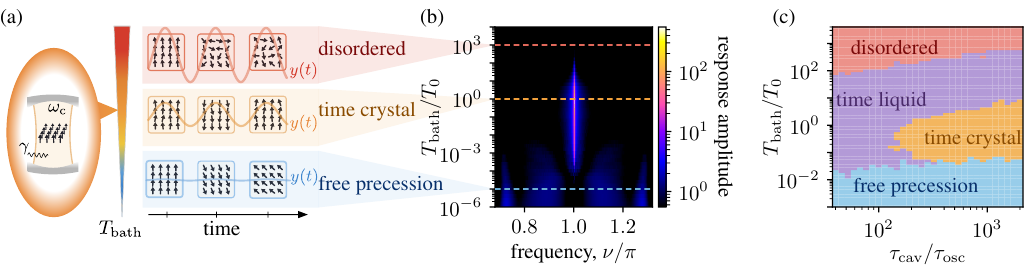}
\caption{
  \textbf{Time liquid in a cavity.}
  (a) Schematic of the setup and three regimes reached by increasing bath temperature. A spin lattice is coupled to a cavity mode of frequency $\omgc = \frac{2 \pi}{\tau_{\rm osc}}$ and damping rate $\gamma = \frac{2}{\tau_{\rm cav}}$, which is in contact with a thermal bath at temperature $\Tbath$ (color bar). The thermal fluctuations of the oscillator coordinate $y(t)$ act on the spins as an effective drive. Each row illustrates the dynamics at the temperature of the corresponding dashed line in (b). At low temperature (blue), the effective drive $y(t)$ is too weak to disturb the spins, which precess freely. At intermediate temperature (orange), the spins lock to a sharp subharmonic response at half the cavity frequency. At high temperature (red), the drive is too strong and the dynamics becomes disordered. Period-doubled order therefore emerges only at intermediate bath temperature. 
  (b) Spin response amplitude $|\tilde F[\llbracket M^z\rrbracket](\nu)|$ as a function of frequency $\nu/\pi$ and bath temperature $\Tbath/T_0$, at $\varepsilon_0=0.1$, $\tosc=0.2$, and $\gamma=10^{-4}$. The subharmonic response at $\nu/\pi=1$ appears within a finite window of bath temperature and vanishes at both small and large $\Tbath$. The period doubling is thus induced by thermal noise. 
  (c) Phase diagram as a function of the cavity coherence time $\tcav/\tosc$ and the bath temperature $\Tbath/T_0$, at $\varepsilon_0=0.05$ and $\tosc=0.01$. We identify four regimes. At low temperature, the spins precess freely because the effective drive is too weak to disturb them. At high temperature, the dynamics is disordered and the period doubling is lost within a few cavity periods. In between, the system forms a DTC or a DTL, according to whether the correlation time $\llbracket\tcorr\rrbracket$ is comparable to or much shorter than the thermalization time $\tth$ (see ``Phase classification'' in Methods). The DTC appears in an intermediate temperature window that's getting narrower as the coherence time decreases and disappears below $\tcav/\tosc\approx130$. The DTL, in contrast, persists at every coherence time shown. A short coherence time of the cavity mode therefore turns the DTC into a DTL.
  Panels (b) and (c) share $L=100$, $J=5$, $M_{\rm osc}=k_B=\lambda=1$, $\theta_{\rm init}=0.2$, $N_p=10^4$ stroboscopic periods, and $N_{\rm traj}=100$ noise realizations. See text for details.
}
\label{fig:fig1}
\end{figure*}

This raises an intriguing question: can temporal order emerge out of noise? To study this question we consider the consequences of coupling a set of spins to a finite-temperature cavity mode (see Fig.~\ref{fig:fig1}). Thermal fluctuations of the cavity generate an approximately periodic signal centered at the cavity frequency. Fluctuations in its amplitude and phase, however, cause this signal to lose temporal coherence beyond a characteristic autocorrelation time. One might therefore expect any induced temporal order to disappear on the same timescale. Instead, we find that the spins enter a long-lived metastable state with a sharp period-doubled response at half the cavity frequency (Fig.~\ref{fig:fig1}(b)). This response persists far beyond the coherence time of the cavity itself (Fig.~\ref{fig:fig3}).

Despite the presence of a sharp subharmonic response, the state lacks the long-range temporal coherence of a time crystal. Instead, it can be viewed as a DTC containing a finite density of time-solitons. Individual trajectories exhibit long intervals of period-doubled dynamics, interrupted by stochastic switches between the two equivalent temporal configurations. We identify these switches as time-solitons: topological defects in temporal order that locally shift the phase of the period-doubled response. Their finite density destroys long-range temporal coherence while preserving local time-crystalline order. We term this defect-rich state a \textit{discrete time liquid} (DTL), in analogy with the melting of a spatial crystal whose long-range order is destroyed by the proliferation of topological defects~\cite{Halperin78}. This picture motivates a minimal stochastic deformation of the DTC evolution in Eq.~\eqref{eq:DTCequation},
\begin{equation}
\hat U_n = e^{i J\sum_{\langle ij\rangle}\hat{S}^z_i\hat{S}^z_j} e^{i(\pi+\varepsilon_n)\sum_i\hat{S}^x_i},\,
\varepsilon_n=
\begin{cases}
\pi & \text{prob. } p\\
0 & \text{prob. } 1-p ,
\end{cases}
\label{eq:DTLequation}
\end{equation}
where $p$ sets the density of time-solitons: with probability $p$, the usual $\pi$ spin flip is replaced by a $2\pi$ rotation, producing a switch between the two temporal configurations. We show that this minimal stochastic-switching model reproduces not only the qualitative phenomenology of the cavity-spin system, but also quantitatively matches the distribution of its temporal correlations.

The DTL regime emerges only once thermal fluctuations become strong enough to switch the interaction and the transverse field on and off out-of-phase within a single cavity cycle. This leads to a threshold temperature for the onset of period-doubled order. Counterintuitively, temporal order therefore appears upon heating, providing another example in which noise promotes order rather than destroying it. Many-body interactions, in turn, stabilize the resulting state, allowing the DTL phase to persist over a broad temperature window, as shown in the phase diagram (Fig.~\ref{fig:fig1}(c)).

\section{Spins coupled to a thermal cavity mode}

We consider a two-dimensional spin lattice coupled to a single damped cavity mode, which we model as a harmonic oscillator in contact with a thermal bath at temperature $\Tbath$ (sketched in Fig.~\ref{fig:fig1}(a)). Crucially, no external periodic drive is applied, so that the thermal fluctuations of the bath are the only time-dependent input to the total system. To access long-time dynamics at large system sizes, we treat the spins classically and consider the spin Hamiltonian:
\begin{align}
H_{\rm spin}(t) &= -J_{\rm eff}(t)\sum_{\langle ij\rangle} S^z_i S^z_j - g_{\rm eff}(t)\sum_i S^x_i, \label{eq:Hspin}\\
J_{\rm eff}(t) = &J\left[1+\lambda y(t)\right], \quad g_{\rm eff}(t) = g_0\left[1-\lambda y(t)\right],\label{eq:couplings}
\end{align}
where $\langle ij\rangle$ denote nearest-neighbor pairs on an $L\times L$ square lattice with periodic boundary conditions, $S^\alpha_i$ is the $\alpha$-component of the classical unit spin at site $i$, $J$ is the bare Ising coupling, $g_0$ is the bare transverse-field strength, $y(t)$ is the cavity oscillator coordinate, and $\lambda$ is the spin--cavity coupling. The cavity coordinate obeys the Langevin equation~\cite{SM}: 
\begin{equation}
\ddot{y} + \gamma \dot{y} + \omega_0^2 y = \xi(t), 
\label{eq:Langevin}
\end{equation}
where $\gamma$ is the damping rate, $\omega_0$ is the natural frequency of the cavity, and $\xi(t)$ is Gaussian white noise with zero mean and correlator $\langle\xi(t)\xi(t')\rangle = D \delta(t-t')$. The noise strength $D = 2\gamma k_B \Tbath/(M_{\rm osc}L^2)$ is fixed by the bath temperature, with $M_{\rm osc}$ the oscillator mass and $L$ the linear system size (see the microscopic derivation in the Supplemental Material (SM)~\cite{SM}). In the underdamped regime $\gamma\ll\omega_0$, the cavity mode acts as a spectral filter for the white noise. As a result, the thermal fluctuations of the bath generate an approximately periodic signal $y(t)$ around the damped frequency $\omgc=\sqrt{\omega_0^2-(\gamma/2)^2}\simeq\omega_0$, with period $\tosc=2\pi/\omgc$. This signal therefore acts on the spins as an effective, noisy periodic drive. Unlike the externally imposed periodic drive of a Floquet DTC~\cite{Else2016PRL,Yao2017PRL,Zaletel2023RMP}, its amplitude is set by the bath temperature, and its phase stays coherent only over a finite coherence time of the cavity mode. Throughout, ``drive'' refers to this effective drive. In the following, we show that it can induce temporal order in the spin lattice.

\begin{figure*}[htbp]
\centering
\includegraphics{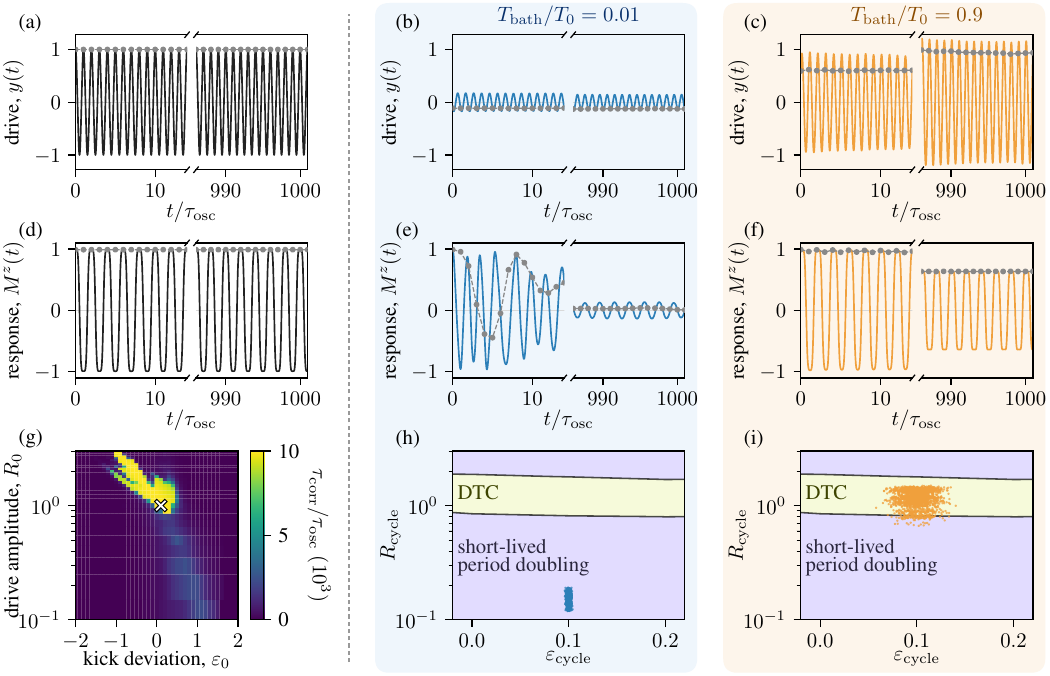}
\caption{
  \textbf{Thermal fluctuations as an effective Floquet drive.}
  Columns compare a deterministic reference drive (a,d,g), in which the oscillator is replaced by a sinusoid, with single realizations of the bath-driven dynamics at a small temperature $\Tbath/T_0=0.01$ (b,e,h, shaded in blue) and at a temperature inside the ordering window $\Tbath/T_0=0.9$ (c,f,i, shaded in orange). 
  (a--c) Oscillator coordinate $y(t)$, which sets the instantaneous interaction and transverse field, with gray dots marking its stroboscopic values. The deterministic drive is a sinusoid of fixed amplitude (a), whereas the stochastic drive is a small (b) or a large (c) fluctuating signal. The bath temperature thus sets the amplitude of the drive. 
  (d--f) Corresponding spin response $M^z(t)$. The deterministic drive gives long-lived period doubling (d). The weak stochastic drive produces a response close to the free precession of the undriven spin lattice, with no period doubling left after $10^3$ periods (e). The large stochastic drive sustains period doubling beyond $10^3$ periods (f). A stochastic drive of sufficient amplitude therefore reproduces the long-lived period doubling of the deterministic drive.
  (g) Correlation time $\tcorr/\tosc$ of Eq.~\eqref{eq:tauDTC_single_traj} for the deterministic drive, over the plane of kick deviation $\varepsilon_0$ and drive amplitude $R_0$. The cross marks the drive involved in (a,d), with $\varepsilon_0=0.1$ and $R_0=1$. Long-lived period doubling appears in a broad region in $(\varepsilon_0,R_0)$, so the DTC does not require fine tuning of the drive amplitude.
  (h,i) Per-cycle effective drive parameters $\{(\varepsilon_{\rm cycle}(n), R_{\rm cycle}(n))\}_n$ of the realizations in (b) and (c) (see ``Per-cycle effective drive parameters'' in Methods), overlaid on the $\tcorr/\tosc=2\times10^3$ contour of (g). This contour separates the DTC band from the surrounding region of short-lived period doubling. The per-cycle effective drive parameters lie below the band at $\Tbath/T_0=0.01$ (h) and fall within it at $\Tbath/T_0=0.9$ (i). Period doubling therefore appears when the per-cycle cloud overlaps the DTC band, which explains the responses in (e) and (f).
  The stochastic columns use $\gamma=10^{-3}$ and $\varepsilon_0=0.1$. All panels share $L=100$, $J=5$, $\tosc=0.2$, $M_{\rm osc}=k_B=\lambda=1$, $\theta_{\rm init}=0.2$, and $N_p=10^4$. See text for details.
}
\label{fig:fig2}
\end{figure*}

\section{Period doubling induced by thermal noise}

We first establish the connection to the standard DTC of Eq.~\eqref{eq:DTCequation}. For times short compared with the coherence time of the mode, $\tcav=2/\gamma$~\cite{SM}, the oscillator coordinate takes the form $y(t)\simeq R\cos(\omgc t+\varphi)$, with a slowly varying amplitude $R$ and phase $\varphi$. The typical amplitude grows with the bath temperature, $\langle R^2\rangle=2\Tbath/T_0$, where $T_0\equiv M_{\rm osc}\omega_0^2L^2/k_B$ is a reference temperature~\cite{SM}. When $R=\lambda=1$, the interaction and the transverse field in Eq.~\eqref{eq:couplings} are switched on out-of-phase. The dynamics over one period then approximately reproduce the kicked evolution of Eq.~\eqref{eq:DTCequation}, the model underlying Floquet DTCs~\cite{Else2016PRL,Yao2017PRL,Zaletel2023RMP}. We thus set $g_0=(\pi+\varepsilon_0)/\tosc$, so that the spins rotate by $\pi+\varepsilon_0$ per period, with a fixed deviation $\varepsilon_0$ from the ideal $\pi$ kick. The bath temperature then controls how closely the thermal fluctuations realize this kicked evolution: a weak drive cannot separate the interaction from the transverse field in time, whereas a strong drive is too erratic. Period-doubled order is therefore expected to emerge only at intermediate bath temperatures.

To test this expectation, we start from the extremely underdamped regime, $\tcav/\tosc=10^5$, which is longer than the observation window of $N_p=10^4$ stroboscopic periods. The phase $\varphi$ is then effectively fixed in each realization, and the amplitude $R$ is the dominant stochastic variable. To quantify the response, we consider the stroboscopic spin magnetization $M^z(n\tosc) = L^{-2}\sum_i S^z_i(n\tosc)$, $n\in \mathbb{Z}$, and the Fourier transform of its noise average, $\tilde F[\llbracket M^z\rrbracket](\nu)$. Here, $\llbracket\cdots\rrbracket$ denotes the average over noise realizations. The transform $\tilde F[f](\nu) = \tosc\sum_{n=0}^{N_p-1} e^{-i\nu n} f(n\tosc)$, is the discrete counterpart of $\int\!dt\,e^{-i\omega t}f(t)$, with $\nu=\omega\tosc$ the dimensionless frequency, so that $\tilde F$ carries units of time. We initialize the lattice close to full polarization along $+z$, tilting each spin by a small random angle within a cone of half-angle $\theta_{\rm init}=0.2$. This tilt mimics the thermal fluctuations of a physical spin system and shows the robustness of the temporal order to different initial conditions. We then solve the coupled dynamics numerically (see ``Initial conditions and simulation parameters'' in Methods). Indeed, the subharmonic response at $\nu = \pi$ emerges within a finite window of bath temperature (see the bright vertical stripe in Fig.~\ref{fig:fig1}(b)). It is absent at low $\Tbath$, where side peaks reflect the precessional dynamics, and it is washed out at high $\Tbath$, where the noise is too strong. More intuitively, at high $\Tbath$ the local field $\vec{B}_i(t)$ inherits a large amplitude from $y(t)$. A tiny spatial difference in the spin configuration then produces a large difference in the local field, which in turn amplifies the spatial difference after a short time. As a result, at high $\Tbath$ the spin lattice develops chaotic dynamics and loses temporal order (top row of Fig.~\ref{fig:fig1}(a)).

We note that the period-doubled response at intermediate bath temperature is absent from the drive, which carries weight only near $\nu = 2\pi$~\cite{SM}. The response is also robust against randomness in the initial spin configuration and against perturbations of the kick deviation $\varepsilon_0$~\cite{SM}. The period-doubled signal is therefore generated by the spin lattice, not inherited from the drive. To understand its origin, we compare the stochastic dynamics with a deterministic reference drive in Fig.~\ref{fig:fig2}.

To this end, we replace the oscillator by the deterministic drive $y(t)=R_0\cos(\omgc t+\phi_0)$ with fixed amplitude $R_0$ and phase $\phi_0$, i.e., the monochromatic limit of the stochastic signal (see Fig.~\ref{fig:fig2}(a)). This reference drive plays the role of the externally imposed drive of a Floquet DTC and is not part of the cavity-spin system. The lifetime of the period doubling depends on $\phi_0$, but this dependence weakens as the drive frequency increases~\cite{SM}. We set $\phi_0=0$, where the lifetime is longest. To measure how long the period doubling persists, we use the normalized staggered magnetization $s(n)\equiv(-1)^nM^z(n\tosc)/M^z(0)$, which stays at $s(n)=1$ for an ideal period-doubled response. We then define the correlation time
\begin{equation}
\tcorr = \tosc\left|\sum_{n \ge 0} s(n)\right|,
\label{eq:tauDTC_single_traj}
\end{equation}
over which a trajectory keeps the memory of its initial subharmonic phase. For an exponential decay, $s(n)=e^{-n\tosc/\tcorr}$ with $\tcorr\gg\tosc$, Eq.~\eqref{eq:tauDTC_single_traj} returns the decay time. Equivalently, $\tcorr=|\tilde F[M^z](\pi)|/|M^z(0)|$ is the height of the subharmonic peak of the Fourier transform used in Fig.~\ref{fig:fig1}(b), normalized by the initial magnetization. Evaluating $\tcorr$ over a grid of $(\varepsilon_0,R_0)$ gives the DTC phase diagram of Fig.~\ref{fig:fig2}(g), in which the bright band marks a long-lived subharmonic response, with typical dynamics shown in Fig.~\ref{fig:fig2}(d). As in prethermal DTCs~\cite{Else2017PRX,Machado2020PRX,Ye2021PRL}, $\tcorr$ grows exponentially with the drive frequency $\omgc$~\cite{SM}.

With the reference phase diagram of Fig.~\ref{fig:fig2}(g) at hand, we now turn to the stochastic drive. At a low temperature, $\Tbath/T_0=0.01$, the drive amplitude is small (Fig.~\ref{fig:fig2}(b)), so that the drive cannot separate the interaction from the transverse field in time. The response then stays close to the free precession of the undriven lattice (Fig.~\ref{fig:fig2}(e)), which gives the side peaks in Fig.~\ref{fig:fig1}(b). 
To compare such a stochastic drive with the deterministic one, we assign to every cavity cycle $n$ an effective kick deviation $\varepsilon_{\rm cycle}(n)$ and an effective drive amplitude $R_{\rm cycle}(n)$ (see ``Per-cycle effective drive parameters'' in Methods). Each realization thereby becomes a cloud of points in the plane of Fig.~\ref{fig:fig2}(g). At low temperature, this cloud lies far below the DTC band (Fig.~\ref{fig:fig2}(h)). Increasing the temperature pushes the cloud upward, since the drive amplitude follows a Rayleigh distribution with $\langle R^2\rangle=2\Tbath/T_0$~\cite{SM}. This distribution is broad, so that the amplitude differs strongly between realizations. The DTC band of Fig.~\ref{fig:fig2}(g), however, is broad in $R_0$ as well, a consequence of the many-body interactions. In an intermediate temperature window the two therefore overlap substantially, and the period doubling survives without any fine tuning of its amplitude. Indeed, at $\Tbath/T_0=0.9$ the drive is a strong fluctuating signal (Fig.~\ref{fig:fig2}(c)) whose cloud sits inside the DTC band (Fig.~\ref{fig:fig2}(i)). The period doubling then persists beyond a thousand cycles (Fig.~\ref{fig:fig2}(f)).

\begin{figure*}[htbp]
\centering
\includegraphics{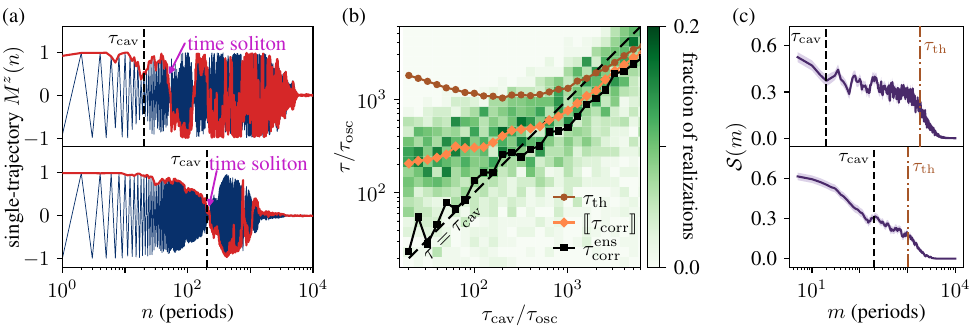}
\caption{
  \textbf{Time-solitons and time liquid lifetime.}
  (a) Single-trajectory stroboscopic magnetization $M^z(n\tosc)$ (blue) and its period-doubled component $(-1)^nM^z(n\tosc)$ (red), at $\tcav/\tosc=20$ (top) and $200$ (bottom). The dashed line marks $\tcav$. The period-doubled component is repeatedly suppressed and revives with the opposite sign. Each such event is a time-soliton, the $\varepsilon_n=\pi$ event of Eq.~\eqref{eq:DTLequation}. 
  (b) Thermalization time $\tth$ (circles), noise averaged single-trajectory correlation time $\llbracket\tcorr\rrbracket$ (diamonds), and ensemble correlation time $\tcorrens$ (squares), as a function of the coherence time $\tcav$. The heatmap gives the fraction of noise realizations per correlation time $\tcorr$ at fixed $\tcav$. The three time scales converge at long $\tcav$, where solitons are rare. They separate as $\tcorrens\leq\llbracket\tcorr\rrbracket\leq\tth$ at short $\tcav$, where solitons proliferate. Solitons thus destroy long-range temporal coherence. 
  (c) Windowed subharmonic weight $\mathcal S(m)$ of Eq.~\eqref{eq:Stm}, measured in a window opened at period $m$, for the two coherence times of (a). The dashed and dash-dotted lines indicate $\tcav$ and $\tth$. Shaded areas mark the standard error over realizations. The weight stays finite in windows opened far beyond $\tcav$ and falls to zero only near $\tth$. Period doubling is therefore re-established after every time-soliton and lost only when the lattice thermalizes.
  All panels share $L=100$, $J=5$, $M_{\rm osc}=k_B=\lambda=1$, $\theta_{\rm init}=0.2$, $N_p=10^4$, $N_{\rm traj}=100$, $\tosc=\varepsilon_0=0.001$, and $\Tbath/T_0=1$. See text for details.
}
\label{fig:fig3}
\end{figure*}

\section{Time-solitons and the discrete time liquid}

We now consider the consequence of a short coherence time of the cavity mode on the temporal order. This finite coherence time has no counterpart in an externally imposed periodic drive. To see this effect, we first make the spin order as robust as possible. Since the lifetime of the period doubling grows exponentially with the cavity frequency~\cite{SM}, we take a high cavity frequency, $\tosc=0.001$, and a small kick deviation, $\varepsilon_0=0.001$. We fix the bath temperature at $\Tbath/T_0=1$, well inside the ordered region of Fig.~\ref{fig:fig2}(g). We then analyze the period-doubled response as a function of the cavity coherence time $\tcav$. Figure~\ref{fig:fig3}(a) shows the single-trajectory dynamics at two coherence times. The period-doubled component $(-1)^nM^z(n\tosc)$ is repeatedly suppressed and revives with the opposite sign. Once the phase of $y(t)$ has drifted by half a period, the lattice rebuilds its subharmonic response against the new phase. The two period-doubled configurations are thereby exchanged. We identify each such event as a time-soliton, the $\varepsilon_n=\pi$ event of Eq.~\eqref{eq:DTLequation}, occurring at a time set by the noise. What limits the temporal order is therefore the proliferation of time-solitons, not the loss of period doubling.

To distinguish different ways of losing the temporal order, we use the normalized staggered magnetization $s(n)$ of Eq.~\eqref{eq:tauDTC_single_traj}, which satisfy $s(0)=1$ in every realization. We introduce the thermalization time and two correlation times before and after ensemble averaging. We define the correlation time of the ensemble-averaged response as: 
\begin{equation}
    \tcorrens \equiv \tosc \left|\sum_{n\ge0} \llbracket s(n)\rrbracket \right| ,\label{eq:tcorrens_def}
\end{equation}
where we average the staggered magnetization first and then compute its correlation time. Then we define the averaged single-trajectory correlation time, $\llbracket \tcorr \rrbracket$, where we compute the correlation time $\tcorr$ of Eq.~\eqref{eq:tauDTC_single_traj} for each individual trajectory, and then average over trajectories. Finally, we define the thermalization time $\tth$ as the overall decay of magnetization irrespective of the sign, which captures the loss of spatial order due to the slow heating,
\begin{equation}
    \tth \equiv 
    \tosc \sum_{n\ge0} \left\llbracket |s(n)| \right\rrbracket . \label{eq:tth_def}
\end{equation}
Here, $\tth$ is insensitive to the subharmonic phase and measures how long the period doubling survives. In contrast, $\llbracket\tcorr\rrbracket$ measures how long one realization spends in its dominant crystalline configuration, and $\tcorrens$ measures how long the ensemble shares a phase. The three obey the inequality $\tcorrens \leq \llbracket\tcorr\rrbracket \leq \tth$, with equality when no soliton occurs within the prethermal window. In the presence of time-solitons, the three time scales separate.

Figure~\ref{fig:fig3}(b) shows the three time scales defined above as a function of the cavity coherence time $\tcav$. We see that the thermalization time depends only weakly on $\tcav$. As in DTCs, the thermalization time is mainly controlled by the cavity frequency. We show its scaling with the cavity frequency in the SM~\cite{SM}. As mentioned earlier, the single-trajectory correlation time $\llbracket\tcorr\rrbracket$ in Fig.~\ref{fig:fig3}(b) lies in between the other two scales. Its distribution over trajectories is shown by the green density plot in Fig.~\ref{fig:fig3}(b). 

\begin{figure}
\centering
\includegraphics[trim= 5 5 0 0, clip]{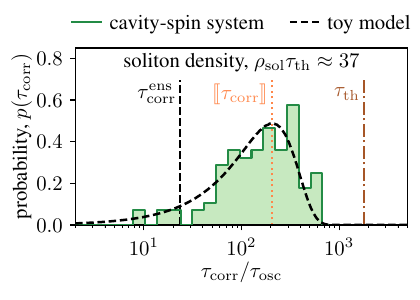}
\caption{
  \textbf{Distribution of the single-trajectory correlation time.}
  Probability $p(\tcorr)$ of the single-trajectory correlation time $\tcorr$ of Eq.~\eqref{eq:tauDTC_single_traj}, for the cavity-spin system (green histogram) and for the toy model of Eqs.~\eqref{eq:classicaltoy} and~\eqref{eq:classicaltoy2} (dashed line), at $\tcav/\tosc=20$. The toy parameters are not fitted: $\tth$ is the value measured on the lattice, and the soliton density $\nsol$ follows from the measured $\tcorrens$, which gives $\nsol\tth\approx37$ (top). The vertical lines mark $\tcorrens$, $\llbracket\tcorr\rrbracket$, and $\tth$ from the line cut at $\tcav/\tosc=20$ in Fig.~\ref{fig:fig3}(b). The toy model reproduces the probability distribution and thus captures the dynamics of the DTL in cavity-spin system. Parameters are as in Fig.~\ref{fig:fig3}. See text for details.
}
\label{fig:fig4}
\end{figure}

To check quantitatively whether this distribution of correlation times can be captured by a simpler model that contains only uncorrelated time-solitons, we propose a classical version of the quantum toy model of Eq.~\eqref{eq:DTLequation}, described as: 
\begin{equation}
    s(n) = e^{-n\tosc/\tth} \times \sigma(n), 
\label{eq:classicaltoy}
\end{equation}
where $\sigma(n)=\pm1$ is a random variable that flips sign with probability $p$ in each period: 
\begin{equation}
    \sigma(n) = \sigma(n-1) \times 
    \begin{cases}
    -1 & \text{prob. } p\\
    +1 & \text{prob. } 1-p .
    \end{cases}
    \label{eq:classicaltoy2}
\end{equation}
Using this model, we show in the Methods that the density of time-solitons, $\nsol$, is determined by the flip probability, $\nsol = p/\tosc$, and that the ensemble correlation time is $\tcorrens = 1/(2\nsol + \tth^{-1})$. Remarkably, despite its simplicity this model reproduces the full probability distribution of the single-trajectory correlation time of the many-body cavity-spin system, as shown in Fig.~\ref{fig:fig4}. To obtain this agreement, we only match $\tth$ and $\tcorrens$ between the toy model and the full simulation (see SM~\cite{SM} for more evidence at other values of $\tcav$). This agreement provides confidence that the time-soliton description captures the dynamics of the DTL. 
Finally, we show in Fig.~\ref{fig:fig3}(b) that the correlation time of the averaged subharmonic response matches the coherence time of the cavity mode, $\tcorrens\simeq\tcav$ in the limit of $\tcav \ll \tth$. This implies that the cavity lifetime directly specifies the density of time-solitons as $\nsol \simeq 1/(2\tcav)$. In the other limit of $\tcav \gg \tth$,  $\tcorrens$ saturates to the thermalization time $\tth$; here, the spatial order is lost before a single time-soliton has time to occur. The system then keeps a single subharmonic phase throughout its lifetime, which is the defining property of a prethermal DTC. The DTC and the DTL are therefore distinguished by the number of time-solitons within the lifetime, $\nsol\tth$.

We further test the time-soliton picture in the cavity-spin system. Equations~\eqref{eq:classicaltoy} and~\eqref{eq:classicaltoy2} predict that a time-soliton affects only the phase of the subharmonic response and leaves its amplitude intact, up to an overall extremely slow thermalization decay. To test this, we measure the local period-doubled order in a window of $W_{\rm cav}=\lfloor\tcav/\tosc\rceil$ periods opened at the stroboscopic period $m$,
\begin{equation}
\mathcal S(m) = \frac{1}{W_{\rm cav}}\left|\sum_{k=0}^{W_{\rm cav}-1}\llbracket s(m+k) s(m)\rrbracket\right|.
\label{eq:Stm}
\end{equation}
Because it averages the product of two magnetizations, $\mathcal S(m)$ never refers to the subharmonic phase of any single realization. It therefore still registers period doubling after the solitons have scrambled that phase across the ensemble. As shown in Fig.~\ref{fig:fig3}(c), the weight $\mathcal S(m)$ stays large in windows opened far beyond $\tcav$ and falls only near the thermalization time $\tth$. At this time, the lattice has absorbed enough energy to melt the spatial order and completely lose the period doubling. Time-solitons only reset the phase of the subharmonic response and leave its amplitude approximately intact for times between $\tcav$ and $\tth$. 

In the phase diagram of Fig.~\ref{fig:fig1}(c), we find that temporal order shows up as a prethermal DTL or as a prethermal DTC, depending on whether the correlation time $\llbracket\tcorr\rrbracket$ is much less than the thermalization time $\tth$ or comparable to it. We therefore use the ratio $\llbracket\tcorr\rrbracket/\tth$ to delineate the phase boundary as shown in the ``Phase classification'' section of the Methods. In that section, we further describe how to capture the boundary to fully chaotic dynamics at high temperatures, and the boundary to purely precessional dynamics at low temperatures. These two more boundaries complete the phase diagram of Fig.~\ref{fig:fig1}(c).

\section{Experimental observability}

In ultracold atoms, our proposal is closely related to optically driven Bose gases coupled to an optical cavity~\cite{Kongkhambut22}. In this geometry, an off-resonant laser typically pumps the Bose--Einstein condensate directly, but the cavity mode can instead be pumped. Driving the cavity directly with white noise at ultralow temperature would then realize our model, with the noise amplitude playing the role of the bath temperature in our calculation.

In condensed matter, a promising platform involves surface phonon polaritons (SPPs), which host ultrastrong electric-field fluctuations~\cite{Emil25,Eckhardt25,Eckhardt24,bretscher26,grunwald26}. Although noise from SPPs involves multiple modes, the fluctuations are laterally coherent over a distance $d$ for a two-dimensional layer located a distance $d$ from the interface. Our model could therefore be realized using a two-dimensional array of dipoles with ferroelectric interactions at this separation. The electric-field fluctuations experienced by the dipoles scale as $\langle E^2 \rangle_{\rm SPP} \sim \frac{\hbar \omega_{\rm SPP}}{\epsilon_0 d^3}\left(1+2n_b\left(\frac{\hbar\omega_{\rm SPP}}{k_bT}\right)\right)\rightarrow\frac{k_bT}{\epsilon_0d^3}$~\cite{Emil25,Eckhardt25,Eckhardt24,bretscher26,grunwald26}, where $E$ is the electric field, $\omega_{\rm SPP}$ the SPP frequency, $\epsilon_0$ the vacuum permittivity, $k_b$ Boltzmann's constant, $T$ temperature, $n_b$ the Bose--Einstein distribution and $\hbar$ the reduced Planck constant. These fluctuations can be comparable to the dipoles' internal dynamics, as required by our model~\cite{Emil25}.

For both platforms, the main challenge is engineering the prescribed coupling of the cavity to both the interaction and kick terms. As in DTC experiments, we expect our model to be one example of a broader class exhibiting DTL order. In particular, we expect that any DTC model featuring a probabilistic switch as described in Eq.~\eqref{eq:DTLequation} would lead to DTL phenomenology.

In both platforms, the main observable is the locking of spin--spin correlation functions to half the cavity frequency as a function of fluctuation amplitude. Cavity-coupled cold gases are particularly attractive because individual trajectories can be imaged, allowing time-solitons to be resolved and their distribution functions measured.

\section{Outlook}

In this article, we demonstrate a new form of temporal order, in which robust subharmonic frequency locking coexists with only short-range temporal coherence. This combination arises from the presence of temporal topological defects, which disrupt long-range coherence but preserve the characteristic subharmonic response.

Our results establish a connection between many-body temporal order and cavity–matter hybrids. Considerable efforts have been devoted to reaching the ultrastrong light–matter coupling regime~\cite{Kipp25}, yet the novel collective phenomena that may emerge in this regime remain largely unexplored. Here, we show that even a relatively simple system of interacting two-dimensional spins coupled to a cavity can exhibit unconventional dynamics governed by temporal order. This suggests that temporal ordering phenomena may play an important role in more complex cavity–matter systems, including electrons in a material lattice coupled ultrastrongly to light.

From the perspective of temporal order, our results point towards a broader landscape of discrete time liquids (DTLs). The fact that the complex spin–cavity dynamics can be captured by a simple toy model governed by a single parameter, the time-soliton density, suggests that the phase studied here may belong to a broader universality class. More generally, DTLs may emerge through different mechanisms and exhibit distinct universality classes, potentially determined by the symmetries and correlations of their temporal defects. Two particularly promising directions for realizing such distinct classes are fluctuating quasiperiodic driving and spin–cavity backaction~\footnote{In this article we neglect the backaction of the spins on the cavity mode, whose effect on the effective drive we discuss in the SM~\cite{SM}. In this case, the time-solitons are set by the cavity fluctuations alone and occur independently. Once the backaction becomes relevant, the cavity mode keeps a memory of the spin dynamics and may generate effective interactions between time-solitons. In the toy model, such interactions can be provided by a flip probability conditioned on the history of the trajectory. Such interactions could lock the solitons at regular separations, as in the case of soliton crystals~\cite{Cole2017NatPhoton}, or enhance their proliferation. They would thereby either strengthen or weaken the temporal coherence.}. In the former, multiple incommensurate cavity frequencies could introduce additional synthetic dimensions~\cite{Martin2017PRX,Else2020PRX}, potentially allowing for higher-dimensional temporal topological defects beyond the isolated time-solitons considered here. In the latter, backaction could mediate interactions between time-solitons, introducing correlations that enhance or suppress temporal coherence. Both mechanisms could thus give rise to DTLs with qualitatively different defect structures and collective dynamics. The possible existence of distinct DTL universality classes further raises the question of whether sharp transitions between temporal phases exhibit universal critical behaviour, analogous to phase transitions in spatially ordered systems.

Finally, the most striking aspect of DTLs, both quantum and classical, is the idea that a DTC with a finite density of topological time defects can restore time-translation symmetry on average. This restoration opens a pathway for temporal order to emerge in thermal equilibrium, thereby circumventing the standard no-go theorems~\cite{Bruno2013PRL,Watanabe2015PRL} that prohibit conventional time crystals in equilibrium in closed isolated systems. At the same time, the resulting phase retains many of the defining features of time-crystalline behavior, effectively bringing temporal order into the realm of equilibrium thermodynamics.

\section*{Methods}

\subsection*{Model and equations of motion}

We numerically simulate the dynamics of the two-dimensional classical spin model with Hamiltonian of Eq.~\eqref{eq:Hspin}, with the oscillator coordinate $y(t)$ generated from the Langevin equation of Eq.~\eqref{eq:Langevin}. The resulting classical spin dynamics obeys
\begin{equation}
\dot{\vec S}_i=\vec S_i\times\vec B_i(t),\quad
\vec B_i(t)=\left(g_{\rm eff}(t),0,J_{\rm eff}(t)\sum_{j\in{\rm nn}(i)}S_j^z\right),
\label{eq:spin_eom}
\end{equation}
where $\vec B_i(t)$ is the local field and ${\rm nn}(i)$ denotes the four nearest neighbors of site $i$ on the periodic square lattice. We use units with $M_{\rm osc}=k_B=1$ and keep $L=100$, $J=5$ and $\lambda=1$ fixed throughout. The backaction of the spins on the oscillator is neglected, so that $y(t)$ can be generated independently of the spin configuration. Consequences of the back-action are discussed in the SM~\cite{SM}.

For the stochastic simulations, we adjust the bare oscillator frequency $\omega_0$ for each damping rate $\gamma$ such that the damped oscillation frequency satisfies
\begin{equation}
\omgc=\sqrt{\omega_0^2-\left(\frac{\gamma}{2}\right)^2}=\frac{2\pi}{\tosc}.
\label{eq:resonant_frequency}
\end{equation}
This choice fixes the oscillation period at $\tosc$ throughout the damping-rate sweeps. We generate an independent realization of $y(t)$ from Eq.~\eqref{eq:Langevin} for each realization of the Gaussian noise. For the deterministic reference calculations behind Fig.~\ref{fig:fig2}, we instead use the sinusoidal signal $y(t)=R_0\cos(\omgc t)$ with fixed amplitude $R_0$.

\subsection*{Numerical integration}

We integrate Eq.~\eqref{eq:spin_eom} by a splitting between the transverse-field and interaction terms of Eq.~\eqref{eq:Hspin}. Within each time step $dt$, every spin is rotated about the $x$ axis by $g_{\rm eff}dt/2$, then about its local $z$ axis by $J_{\rm eff}\sum_{j\in{\rm nn}(i)}S^z_j dt$, and then about the $x$ axis by $g_{\rm eff}dt/2$ again. The Langevin equation, Eq.~\eqref{eq:Langevin}, is propagated on the same grid, with the noise sampled as $\xi_n=\sqrt{D/dt} \eta_n$, $\eta_n\sim\mathcal N(0,1)$, and held constant within the step. The noise is generated independently at each step and for every noise realization. With small enough $dt$, this trotterization method gives converged dynamics as we show in the SM~\cite{SM}.

\subsection*{Initial conditions and simulation parameters}

We initialize each spin independently and uniformly within a spherical cap of half-angle $\theta_{\rm init}$ about the $+z$ axis:
\begin{equation}
\vec S_i(0)=
\left(
\sin\theta_i\cos\phi_i,
\sin\theta_i\sin\phi_i,
\cos\theta_i
\right),
\label{eq:spin_initialization}
\end{equation}
where $\phi_i\sim U[0,2\pi)$ and $\cos\theta_i\sim U[\cos\theta_{\rm init},1]$. The cap half-angle $\theta_{\rm init}$ sets how far the initial state departs from the perfectly polarized configuration, which is recovered at $\theta_{\rm init}=0$. We use $\theta_{\rm init}=0.2$ in each figure of the main text, so that the lattice starts close to the ordered configuration of the Ising interaction. Before initiating the spin dynamics, we evolve the oscillator from $y=\dot y=0$ for a time $5/\gamma$ to approach its stationary state.

We resolve the cavity period $\tosc$ with at least $20$ time steps, so that $dt=5\times10^{-3}$ at $\tosc=0.2$, $dt=5\times10^{-4}$ at $\tosc=0.01$ and $dt=5\times10^{-5}$ at $\tosc=10^{-3}$. We record $N_p=10^4$ periods $\tosc$ in every run. Noise averages are taken over $N_{\rm traj}=100$ independent noise realizations.

\subsection*{Per-cycle effective drive parameters}

To compare a stochastic drive with the deterministic drive $y(t)=R_0\cos(\omgc t)$, we assign to cycle $n$ the effective kick deviation and amplitude
\begin{align}
\varepsilon_{\rm cycle}(n) &= \varepsilon_0 - g_0\lambda\int_{n\tosc}^{(n+1)\tosc} y(t)\,dt, \label{eq:eps_cycle}\\
R_{\rm cycle}(n) &= \sqrt{\frac{2}{\tosc}\int_{n\tosc}^{(n+1)\tosc} y(t)^2\,dt}. \label{eq:r_cycle}
\end{align}
Here, $\varepsilon_{\rm cycle}(n)$ is the deviation of the spin rotation accumulated in cycle $n$ from $\pi$. The amplitude $R_{\rm cycle}(n)$ is the root-mean-square of $y(t)$ over the cycle, normalized such that $R_{\rm cycle}=R$ for a pure tone $y=R\cos(\omgc t+\varphi)$. A deterministic drive therefore maps to the single point $(\varepsilon_0,R_0)$ in every cycle. For the stochastic drive, $\langle y\rangle=0$ gives $\langle\varepsilon_{\rm cycle}\rangle=\varepsilon_0$. In the underdamped limit, $R_{\rm cycle}\simeq R$ follows the Rayleigh distribution of the drive amplitude. We analytically derive this result in the SM~\cite{SM}.

\subsection*{Toy model of time-solitons}

In the toy model of Eqs.~\eqref{eq:classicaltoy} and~\eqref{eq:classicaltoy2}, a time-soliton occurs independently in each period with probability $p$. Neighboring solitons are therefore separated by $1/p$ periods on average, and the density of time-solitons is $\nsol=p/\tosc$. 
In the limit of $\tth\gg\tosc$ and $\nsol\tosc\ll1$, Eq.~\eqref{eq:tcorrens_def} gives
\begin{equation}
\tcorrens=\frac{1}{2\nsol+\tth^{-1}}.
\end{equation}
We fix $p$ by this relation with the values of $\tth$ and $\tcorrens$ measured in the cavity-spin system, so that the toy model has no free parameter.

The single-trajectory correlation time is $\tcorr=|Z|$, with $Z\equiv\tosc\sum_{k\ge0}a^k\sigma(k)=\tosc(1+2W)/(1-a)$, $a\equiv e^{-\tosc/\tth}$, and $W\equiv\sum_{i\ge1}(-1)^ia^{k_i}$. Here, $k_1<k_2<\dots$ are the periods at which solitons occur. 
Since $W=-a^{k_1}(1+W')$, with $W'$ sharing the same probability distribution with $W$, the cumulative probability distribution $P(W\le w)$ obeys
\begin{equation}
P(W\le w) = \sum_{g\ge1}(1-p)^{g-1}p \left[1-P(W\le -wa^{-g}-1)\right].
\label{eq:toy_cdf_methods}
\end{equation}
We solve Eq.~\eqref{eq:toy_cdf_methods} iteratively and construct the probability distribution of $Z$ and $\tcorr$ (more details are found in the SM~\cite{SM}).

\subsection*{Phase classification}

Each point in the map of Fig.~\ref{fig:fig1}(c) is classified by the correlation time $\llbracket\tcorr\rrbracket$ of Eq.~\eqref{eq:tauDTC_single_traj}, the thermalization time $\tth$ of Eq.~\eqref{eq:tth_def}, and the precession weight
\begin{equation}
P_{\rm prec}=\max\bigl(0,-\min_k \mathcal C(k)\bigr),
\quad
\mathcal C(k)\equiv\frac{\sum_{m}\llbracket s(m+k) s(m)\rrbracket}{\sum_{m}\llbracket s(m)^2\rrbracket},
\label{eq:Pprec}
\end{equation}
where $m$ runs over a reference window of $100$ periods starting at $m=50$. The weight $P_{\rm prec}$ approaches unity when $\mathcal C(k)$ oscillates down to $-1$ and vanishes otherwise. We first identify the boundary of free precession as the lowest temperature at which $P_{\rm prec}$ has fallen to its high-temperature floor. We then label the remaining points as disordered when $\llbracket\tcorr\rrbracket<10 \tosc$. All of the remaining points show local temporal order. We label them a DTC if $\llbracket\tcorr\rrbracket/\tth$ exceeds $0.418$, and a DTL otherwise. More details on the phase classification approach can be found in the SM~\cite{SM}.

\vspace{1cm}

\emph{Data availability.---}%
The data and code associated with this manuscript are deposited in DOI: \href{https://doi.org/10.5281/zenodo.23168407}{10.5281/zenodo.23168407}.
Automatic access is restricted; access credentials will be provided to human researchers upon reasonable request via Zenodo for academic and verification purposes.

\emph{Acknowledgments.---}%
We thank Duillio De Santis, Hope Bretcher and Sambuddha chattopadhyay for valuable discussions.
The authors used Claude Opus 5.5 during the preparation of this manuscript to assist with code optimization and to improve the readability of the manuscript. The authors assume full responsibility for the scientific content, accuracy, and integrity of this paper.

Funded by the Deutsche Forschungsgemeinschaft (DFG, German Research Foundation) – Project number 579633640 (Emmy Noether Programme, “Light–Matter Hybrids”). Funded by the European Union (ERC, QuSimCtrl, 101113633). Views and opinions expressed are, however, those of the authors only and do not necessarily reflect those of the European Union or the European Research Council Executive Agency. Neither the European Union nor the granting authority can be held responsible for them. Z.W. acknowledges financial support from the China Scholarship Council (CSC, No. 202506750006) and the NKRDPC (Grant No. 2022YFA1405304).

\emph{Author contributions.---}
M.H.M.~conceived the research with M.B.;
Z.W.~set up the theoretical framework and performed the numerical simulations with support from P.M.S., M.B., and M.H.M. M.H.M. supervised the project with support from M.B. 
All authors contributed to interpreting the data. Z.W. and M.H.M. wrote the manuscript with input from P.M.S. and M.B.

\emph{Competing Interests Statement.---}
The authors declare no competing interests.

\resumetoc
\ifsuppmode

\cleardoublepage

\renewcommand{\thefigure}{S\arabic{figure}}
\setcounter{figure}{0}
\renewcommand{\theequation}{S.\arabic{equation}}
\setcounter{equation}{0}
\renewcommand{\thesection}{S\arabic{section}}
\setcounter{section}{0}

\onecolumngrid
\begin{center}
{\large\textbf{\textit{Supplemental Material}}}\\
	\hfill \break
\end{center}
\twocolumngrid

\ifcombinedmode
\else
\maketitle
\fi 

\tableofcontents

This Supplemental Material collects the derivations and diagnostics behind the main-text results. Sections~\ref{sm:langevin} and~\ref{sm:rayleigh} derive oscillator Langevin equation, Eq.~\eqref{eq:Langevin}, by eliminating the bath and give the stationary statistics of the filtered drive. Sections~\ref{sm:robustness} and~\ref{sm:spectra} show the robustness of the response and compare the drive and response spectra behind Fig.~\ref{fig:fig1}. Sections~\ref{sm:prethermal} and~\ref{sm:liquid_prethermal} give the prethermal lifetime scaling at long and at short coherence times of the cavity mode. Section~\ref{sm:phase} shows how the drive phase affects the lifetime of the deterministic DTC of Fig.~\ref{fig:fig2}. Section~\ref{sm:toymodel} derives the time scales of the toy model of Eqs.~\eqref{eq:classicaltoy} and~\eqref{eq:classicaltoy2} and the distribution of single-trajectory correlations times behind Fig.~\ref{fig:fig4}. Section~\ref{sm:classification} gives the numerical details and checks behind the phase diagram of Fig.~\ref{fig:fig1}(c). Sections~\ref{sm:backaction} and~\ref{sm:numerics} discuss the spin backaction on the oscillator and the numerical methods.

\section{Langevin equation from thermal bath}\label{sm:langevin}

We derive the oscillator Langevin equation by eliminating the bath, following the standard treatment of a coordinate coupled to a harmonic reservoir~\cite{Ford1965JMP}.

\subsection{Bath model and Langevin equation}

The reservoir is a set of harmonic modes (coordinates $X_\alpha$, momenta $P_\alpha$, masses $m_\alpha$, frequencies $\omega_\alpha$) coupled linearly to $x$ with strengths $c_\alpha$, plus a counterterm cancelling the bath-induced frequency shift,
\begin{align}
H_{\rm bath} &= \sum_\alpha \left[\frac{P_\alpha^2}{2m_\alpha}
+ \frac{1}{2} m_\alpha\omega_\alpha^2 X_\alpha^2\right],\\
H_{\rm osc\text{-}bath} &= -x\sum_\alpha c_\alpha X_\alpha
+ x^2\sum_\alpha \frac{c_\alpha^2}{2m_\alpha\omega_\alpha^2}.
\end{align}
The bare oscillator is $H_{\rm osc} = p^2/2M_{\rm osc} + \frac{1}{2} M_{\rm osc}\omega_0^2 x^2$ ($p$ conjugate to $x$, mass $M_{\rm osc}$, natural frequency $\omega_0$), coupled to the spins through $\lambda\, y\, H^{(-)}$: here $y\equiv x/L$ is the intensive coordinate normalized in Sec.~\ref{sm:thermo}, $\lambda$ the coupling strength, and $H^{(-)}$ (Eq.~\eqref{eq:sm_Hminus}) the extensive spin combination modulating $J_{\rm eff}$ and $g_{\rm eff}$. Hamilton's equations give
\begin{align}
M_{\rm osc}\ddot x + M_{\rm osc}\omega_0^2 x
+ x\sum_\alpha\frac{c_\alpha^2}{m_\alpha\omega_\alpha^2}
&= -\frac{\lambda}{L} H^{(-)}(t) + \sum_\alpha c_\alpha X_\alpha,
\label{eq:sm_eomx}\\
m_\alpha\ddot X_\alpha + m_\alpha\omega_\alpha^2 X_\alpha &= c_\alpha x.
\label{eq:sm_eomX}
\end{align}
Each bath mode~\eqref{eq:sm_eomX} is driven by $x(t)$; integrating from an initial time $t_i$,
\begin{equation}
X_\alpha(t) = X_\alpha^{\rm (h)}(t)
+ \frac{c_\alpha}{m_\alpha\omega_\alpha}\int_{t_i}^{t}\!dt'\,
\sin[\omega_\alpha(t-t')]\,x(t'),
\label{eq:sm_Xsol}
\end{equation}
with the free evolution of mode $\alpha$,
\begin{equation}
X_\alpha^{\rm (h)}(t) = X_\alpha(t_i)\cos[\omega_\alpha(t-t_i)]
+ \frac{P_\alpha(t_i)}{m_\alpha\omega_\alpha}\sin[\omega_\alpha(t-t_i)],
\label{eq:sm_Xhom}
\end{equation}
set by the initial data $X_\alpha(t_i), P_\alpha(t_i)$. Inserting Eq.~\eqref{eq:sm_Xsol} into Eq.~\eqref{eq:sm_eomx} and integrating the memory term by parts, the instantaneous piece $\propto x(t)$ cancels the counterterm, leaving the generalized Langevin equation
\begin{equation}
M_{\rm osc}\ddot x + M_{\rm osc}\omega_0^2 x
+ \int_{t_i}^t\!dt'\,\Gamma(t-t')\,\dot x(t')
= \Xi(t) - \frac{\lambda}{L} H^{(-)}(t),
\label{eq:sm_genlangevin}
\end{equation}
with kernel $\Gamma(t) = \sum_\alpha (c_\alpha^2/m_\alpha\omega_\alpha^2)\cos(\omega_\alpha t)$ and bath force $\Xi(t) = \sum_\alpha c_\alpha X_\alpha^{\rm (h)}(t)$. A boundary term $-x(t_i)\,\Gamma(t-t_i)$ has been dropped ($x(t_i)=0$).

\subsection{Markovian limit and fluctuation--dissipation}

When the bath correlation time is short, $\dot x$ is nearly constant over the memory range of $\Gamma$ and the integral becomes local,
\begin{equation}
\int_{t_i}^t\! dt'\,\Gamma(t-t')\,\dot x(t')
\simeq \dot x(t)\int_0^\infty\! ds\,\Gamma(s)
\equiv M_{\rm osc}\gamma\,\dot x(t),
\end{equation}
defining the damping rate $\gamma$; equivalently $\Gamma(t)\to 2M_{\rm osc}\gamma\,\delta(t)$, the factor of two from the one-sided integral ($\int_0^\infty\! ds\,\delta(s)=\tfrac12$). Equation~\eqref{eq:sm_genlangevin} becomes
\begin{equation}
\ddot x + \gamma\dot x + \omega_0^2 x
= \frac{\Xi(t)}{M_{\rm osc}} - \frac{\lambda}{M_{\rm osc}L}H^{(-)}(t).
\label{eq:sm_langevin_x}
\end{equation}
In thermal equilibrium at temperature $\Tbath$ the modes are independent, with $\langle X_\alpha(t_i)^2\rangle = k_B \Tbath/(m_\alpha\omega_\alpha^2)$, $\langle P_\alpha(t_i)^2\rangle = m_\alpha k_B \Tbath$, and zero means and cross-correlation ($k_B$ Boltzmann's constant). Then $\Xi$ is zero-mean Gaussian and, from Eq.~\eqref{eq:sm_Xhom}, obeys the classical fluctuation--dissipation relation $\langle\Xi(t)\Xi(t')\rangle = k_B \Tbath\,\Gamma(t-t')$. In the Markovian limit it is white,
\begin{equation}
\langle\Xi(t)\Xi(t')\rangle = 2M_{\rm osc}\gamma\, k_B \Tbath\,\delta(t-t').
\label{eq:sm_markov_noise}
\end{equation}
Equations~\eqref{eq:sm_langevin_x}--\eqref{eq:sm_markov_noise} govern the extensive coordinate $x$; its intensive form follows in Sec.~\ref{sm:thermo}.

\subsection{Thermodynamic limit}
\label{sm:thermo}

The oscillator is a single cavity mode, $x\propto \hat a+\hat a^\dagger$, whereas the spin combination it couples to,
\begin{equation}
H^{(-)} = -\sum_{\langle ij\rangle} J\, S^z_i S^z_j + g_0\sum_i S^x_i,
\label{eq:sm_Hminus}
\end{equation}
the part of $H_{\rm spin}$ (Eq.~\eqref{eq:Hspin}) multiplying $x$, is extensive: $H^{(-)}=\mathcal{O}(N)$ ($N=L^2$ spins). At fixed photon density $\langle \hat a^\dagger\hat a\rangle/N=\mathcal{O}(1)$ the mode amplitude scales as $x\sim\sqrt{\langle \hat a^\dagger\hat a\rangle}\sim L$, so we use the intensive coordinate and per-spin force
\begin{equation}
y \equiv \frac{x}{L},
\qquad
h^-(t) \equiv \frac{H^{(-)}(t)}{N},
\end{equation}
both $\mathcal{O}(1)$, with $h^-(t)$ the instantaneous configuration. The coupling $\lambda\, y\, H^{(-)}=\mathcal{O}(N)$ is then extensive while the modulation $\lambda y$ stays $\mathcal{O}(1)$. Dividing Eq.~\eqref{eq:sm_langevin_x} by $L$ (with $H^{(-)}=L^2 h^-$) gives
\begin{equation}
\ddot y + \gamma\dot y + \omega_0^2 y = \frac{\Xi(t)}{M_{\rm osc}L} - \frac{\lambda}{M_{\rm osc}}\,h^-(t),
\end{equation}
with $\mathcal{O}(1)$ backaction. The intensive noise $\xi\equiv\Xi/(M_{\rm osc}L)$ is white by Eq.~\eqref{eq:sm_markov_noise}, the $1/L$ rescaling its strength by $1/L^2$:
\begin{equation}
\langle\xi(t)\xi(t')\rangle = D\,\delta(t-t'),
\qquad
D = \frac{2\gamma k_B \Tbath}{M_{\rm osc}L^2},
\label{eq:sm_FDT}
\end{equation}
the noise strength below Eq.~\eqref{eq:Langevin}. Equivalently, the drive amplitude $\sigma_R^2=k_B\Tbath/(M_{\rm osc}\omega_0^2L^2)=\Tbath/T_0$ is tuned in the main text, with the reference temperature $T_0\equiv M_{\rm osc}\omega_0^2L^2/k_B$ used on the figure axes. Dropping the backaction recovers Eq.~\eqref{eq:Langevin}, $\ddot y+\gamma\dot y+\omega_0^2 y=\xi(t)$.

\section{Stationary statistics of the filtered drive}
\label{sm:rayleigh}

We derive the Rayleigh-amplitude, uniform-phase form of $y(t)$. Without backaction Eq.~\eqref{eq:Langevin} is linear; with $y(t_i)=\dot y(t_i)=0$,
\begin{equation}
y(t) = \int_{t_i}^t\!dt'\,G(t-t')\,\xi(t'),\quad
G(t) = \frac{e^{-\gamma t/2}}{\omgc}\sin(\omgc t)\,\Theta(t),
\end{equation}
with $\omgc = \sqrt{\omega_0^2-\gamma^2/4}$. For $\gamma\ll\omega_0$, coarse-grained over $\omega_0^{-1}\ll t\ll\gamma^{-1}$, we write $y$ in quadratures,
\begin{equation}
y(t) = C(t)\cos\omgc t + S(t)\sin\omgc t,
\end{equation}
where $C, S$ are slowly varying (scale $\gamma^{-1}$) Gaussian processes. In the stationary limit $t-t_i\gg\gamma^{-1}$ they are independent, zero-mean, and of equal variance,
\begin{equation}
\langle C^2\rangle = \langle S^2\rangle = \sigma_R^2,
\qquad \langle CS\rangle = 0,
\end{equation}
with
\begin{equation}
\sigma_R^2 = \frac{D}{2\gamma\omega_0^2}
= \frac{k_B \Tbath}{M_{\rm osc}\omega_0^2 L^2},
\end{equation}
using Eq.~\eqref{eq:sm_FDT}. This variance is exact, fixed by equipartition: $\langle x^2\rangle = k_B \Tbath/(M_{\rm osc}\omega_0^2)$ and $\langle y^2\rangle = \langle x^2\rangle/L^2$; $\omgc\simeq\omega_0$ sets the oscillation, not the amplitude scale. With $C = R\cos\varphi$, $S = -R\sin\varphi$,
\begin{equation}
y(t) \simeq R\cos(\omgc t + \varphi),
\end{equation}
with $\varphi$ uniform on $[0,2\pi)$ and the amplitude $R = \sqrt{C^2+S^2}$ Rayleigh-distributed,
\begin{equation}
f_R(R) = \frac{R}{\sigma_R^2}\,e^{-R^2/2\sigma_R^2}.
\end{equation}
The drive strength $\langle R^2\rangle = 2\sigma_R^2 \propto \Tbath$ thus vanishes as $\Tbath\to 0$ and grows without bound at large $\Tbath$. These two limits bounding the ordering window of Fig.~\ref{fig:fig1}(b). Over times $\ll\gamma^{-1}$, $R$ and $\varphi$ are frozen, so the spins see a quasi-monochromatic drive of random amplitude and phase.

Since the Green function of Eq.~\eqref{eq:Langevin} carries the envelope $e^{-\gamma t/2}$, the stationary correlation of the mode decays as $\langle y(t)y(0)\rangle\propto e^{-\gamma t/2}\cos(\omgc t)$. The time that the mode keeps the memory of its phase is therefore 
\begin{equation}
\tcav = \frac{2}{\gamma} ,
\label{eq:sm_tcav}
\end{equation}
which we call coherence time in the main text. 

\section{Robustness of the noise-induced temporal order}\label{sm:robustness}

\begin{figure}
\centering
\includegraphics{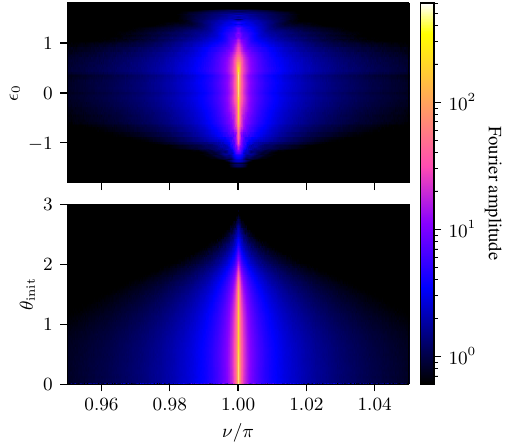}
\caption{
  \textbf{Robustness of the noise-induced period doubling.}
  Fourier amplitude $|\tilde F[\llbracket M^z\rrbracket](\nu)|$ of the noise-averaged response (the Fig.~\ref{fig:fig1}(b) observable) against $\nu/\pi$, at the optimal bath temperature $\Tbath/T_0=1$, as (a)~the static kick deviation $\varepsilon_0$ and (b)~the initial-state randomness $\theta_{\rm init}$ are varied. The subharmonic ridge at $\nu/\pi=1$ persists over a finite band of $\varepsilon_0$ and across a large range of initial-state disorder. The period doubling is therefore robust to both drive-parameter and initial-configuration perturbations. Parameters: $L=100$, $J=5$, $\tosc=0.2$, $\gamma=10^{-4}$, $N_p=10^4$, and $N_{\rm traj}=100$. Panel (a) fixes $\theta_{\rm init}=0.2$, and panel (b) fixes $\varepsilon_0=0.1$.
}
\label{fig:sm_robustness}
\end{figure}

The main text states that the period-doubled response of Fig.~\ref{fig:fig1}(b) is robust against randomness in the initial spin configuration and against perturbations of the kick deviation $\varepsilon_0$. We verify both claims by repeating the Fig.~\ref{fig:fig1}(b) measurement at the optimal bath temperature $\Tbath/T_0=1$ while separately scanning $\varepsilon_0$ and the initial-state cap half-angle $\theta_{\rm init}$. Figure~\ref{fig:sm_robustness} shows the resulting subharmonic response. The bright ridge at $\nu/\pi=1$ survives over a finite window of the kick deviation (Fig.~\ref{fig:sm_robustness}(a)). It also persists across a large swept range of initial-state randomness, decaying only gradually as $\theta_{\rm init}$ approaches the fully random cap $\theta_{\rm init}=\pi$ (Fig.~\ref{fig:sm_robustness}(b)). The noise-induced DTC therefore requires neither fine tuning of the effective kick nor a specially prepared initial state, consistent with the prethermal-DTC mechanism of Fig.~\ref{fig:fig2}.

\section{Comparison between cavity response and period-doubled spin response}
\label{sm:spectra}

\begin{figure}
\centering
\includegraphics{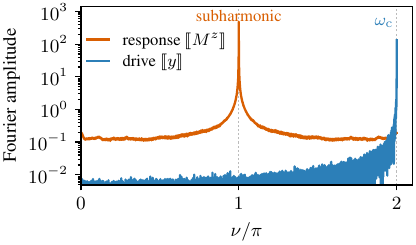}
\caption{
  \textbf{Drive versus response spectra.} Fourier amplitude $|\tilde F[\cdot](\nu)|$ of the noise-averaged drive $\llbracket y\rrbracket$ (blue) and response $\llbracket M^z\rrbracket$ (orange) against $\nu/\pi$ ($\nu=\omega\tosc$), with the parameters of Fig.~\ref{fig:fig1}(b) at $\Tbath/T_0=1$. The drive carries weight at its frequency $\omgc$ ($\nu/\pi=2$) but none at $\nu/\pi=1$. The response develops a sharp subharmonic peak at $\nu/\pi=1$, half the cavity frequency, which is absent from the drive.
}
\label{fig:sm_spectra}
\end{figure}

The period-doubled signal is generated by the spin lattice, not inherited from the drive. Using the same Fourier amplitude of the noise-averaged signal, $|\tilde F[\cdot]|$ (the Fig.~\ref{fig:fig1}(b) observable), Fig.~\ref{fig:sm_spectra} compares drive and response. The drive has weight only near its frequency $\omgc$ ($\nu/\pi=2$), whereas the response develops a sharp subharmonic at $\nu/\pi=1$ that is absent from the drive. The full frequency--temperature dependence of this subharmonic, $|\tilde F[\llbracket M^z\rrbracket](\nu)|$ over $(\nu/\pi, \Tbath)$, is the emergence map of Fig.~\ref{fig:fig1}(b), whose bright ridge at $\nu/\pi=1$ is the period-doubled response.

\section{Prethermal lifetime scaling}\label{sm:prethermal}

\begin{figure}
\centering
\includegraphics{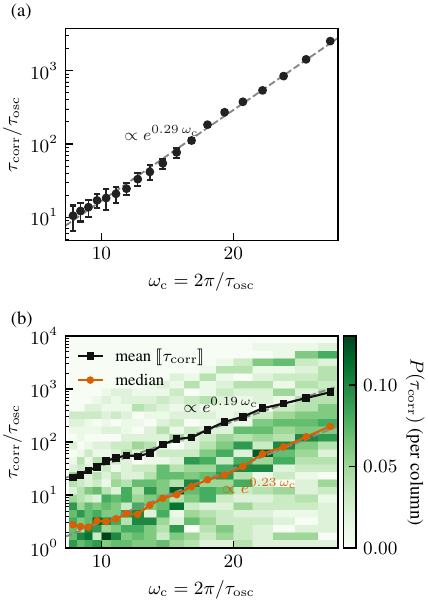}
\caption{
  \textbf{Prethermal lifetime scaling.}
  Correlation time in units of the period $\tosc$, $\tcorr/\tosc$ (Eq.~\eqref{eq:tauDTC_single_traj}), against the frequency $\omgc=2\pi/\tosc$, with dashed exponential guides $\propto e^{c\omgc}$.
  (a)~Deterministic drive at a representative point of the Fig.~\ref{fig:fig2}(g) DTC region ($\varepsilon_0=0$, $R_0=1$), averaged over initial-state realizations (error bars: standard deviation). The lifetime grows exponentially by two orders of magnitude, with $c\simeq0.29$.
  (b)~Stochastic drive at fixed coherence time $\tcav/\tosc=2\times10^5$. The green density shows the per-realization distribution $P(\tcorr/\tosc)$ at each frequency, with the noise-averaged mean $\llbracket\tcorr\rrbracket$ (black squares) and the median (orange circles). Both grow exponentially, with $c\simeq0.19$ and $0.23$.
  Parameters: $L=100$, $J=5$, $\lambda=1$, and $\theta_{\rm init}=0.2$. Panel (b) uses $\Tbath/T_0=1$ and $N_{\rm traj}=100$.
}
\label{fig:sm_prethermal}
\end{figure}

As in prethermal discrete time crystals~\cite{Else2017PRX,Machado2020PRX,Ye2021PRL}, the lifetime of the noise-induced period doubling grows exponentially with the cavity frequency. We test this scaling by fixing the operating point and sweeping the period $\tosc$, hence $\omgc=2\pi/\tosc$. For a deterministic drive at a representative point of the DTC band of Fig.~\ref{fig:fig2}(g) ($\varepsilon_0=0$, $R_0=1$), the correlation time $\tcorr$ (Eq.~\eqref{eq:tauDTC_single_traj}), averaged over initial spin configurations, grows exponentially by two orders of magnitude, $\tcorr/\tosc\propto e^{c\omgc}$ with $c\simeq0.29$ (Fig.~\ref{fig:sm_prethermal}(a)). This is the exponential prethermal scaling invoked in the main text for both Figs.~\ref{fig:fig2} and~\ref{fig:fig3}.

The same scaling survives the stochastic drive. We hold the coherence time of the cavity mode fixed in units of cavity cycles, $\tcav/\tosc=2\times10^5$, so that every frequency is compared at equal quality factor. We then record the single-trajectory correlation time of each noise realization. Figure~\ref{fig:sm_prethermal}(b) shows the resulting distribution against $\omgc$. A few long-lived realizations pull the noise-averaged mean $\llbracket\tcorr\rrbracket$ well above the typical (median) value, Nevertheless, both the mean and the median grow exponentially, with rates $c\simeq0.19$ and $0.23$ comparable to the deterministic value. The prethermal protection generated by the narrowband stochastic drive therefore obeys the same exponential frequency scaling as its deterministic counterpart.

\section{Thermalization time at short coherence time}
\label{sm:liquid_prethermal}

\begin{figure}
\centering
\includegraphics{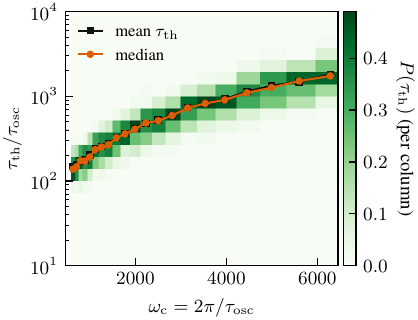}
\caption{
  \textbf{Thermalization time at short coherence time.}
  Thermalization time, $\tth/\tosc$ (Eq.~\eqref{eq:tth_def}), as a function of the cavity frequency $\omgc=2\pi/\tosc$, at the fixed coherence time $\tcav/\tosc=20$ of the DTL in Fig.~\ref{fig:fig4}. The green density shows the distribution of the single-realization value $\tosc\sum_{n}|s(n)|$ at each frequency, with its mean $\tth$ (black squares) and its median (orange circles). The cavity frequency controls the lifetime of the period doubling also in the DTL. Parameters: $L=100$, $J=5$, $\lambda=1$, $\varepsilon_0=0.001$, $\theta_{\rm init}=0.2$, $\Tbath/T_0=1$, $N_p=10^4$, and $N_{\rm traj}=100$.
}
\label{fig:sm_liquid_prethermal}
\end{figure}

Section~\ref{sm:prethermal} shows the exponential growth of the correlation time at a long coherence time, where no time-soliton occurs. Here, we test it in the DTL regime of Fig.~\ref{fig:fig3}, where time-solitons separate the correlation times from $\tth$. We fix the coherence time at $\tcav/\tosc=20$ and sweep the cavity period from $\tosc=0.01$ to $0.001$, i.e., $\omgc$ from $200\pi$ to $2000\pi$, with all other parameters as in Fig.~\ref{fig:fig3}. For each noise realization, we evaluate $\tosc\sum_{n=0}^{N_p}|s(n)|$, whose noise average is the thermalization time $\tth$ of Eq.~\eqref{eq:tth_def}. Figure~\ref{fig:sm_liquid_prethermal} shows that $\tth/\tosc$ grows monotonically, where the distribution at each frequency is narrow. The growth is slower than a single exponential in $\omgc$ but still shows a strong frequency dependence. 

\section{Drive-phase dependence of the DTC lifetime}
\label{sm:phase}

\begin{figure}
\centering
\includegraphics{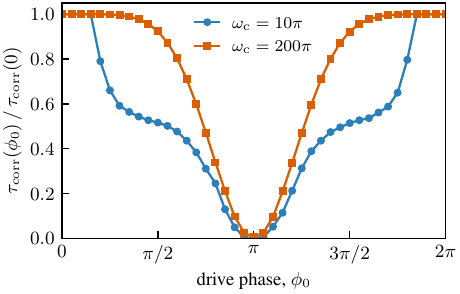}
\caption{
  \textbf{Drive-phase dependence of the DTC lifetime.}
  Correlation time $\tcorr$ of Eq.~\eqref{eq:tauDTC_single_traj}, normalized to its value at $\phi_0=0$, of the deterministic drive $y(t)=R_0\cos(\omgc t+\phi_0)$ as a function of the initial phase $\phi_0$, for a slow ($\omgc=10\pi$) and a fast ($\omgc=200\pi$) drive. At both frequencies, $\tcorr$ is largest near $\phi_0=0$ and vanishes near $\phi_0=\pi$. The drive phase therefore strongly affects the DTC lifetime. As $\omgc$ increases, the plateau of maximal $\tcorr$ around $\phi_0=0$ widens. The phase dependence thus weakens at higher frequency. Parameters: $L=100$, $J=5$, $\varepsilon_0=0$, $R_0=1$, $\theta_{\rm init}=0.2$, $N_p=10^4$.
}
\label{fig:sm_phase}
\end{figure}

To understand the mechanism, the main text replaces the oscillator with a single deterministic tone $y(t)=R_0\cos(\omgc t+\phi_0)$ and fixes the phase to $\phi_0=0$ (Fig.~\ref{fig:fig2}(a,d,g)). The full stochastic dynamics (Fig.~\ref{fig:fig2}(b,c,e,f,h,i)), however, sees a filtered drive $y(t)\simeq R\cos(\omgc t+\varphi)$. Its phase $\varphi$ is uniform on $[0,2\pi)$ and takes a different value in each noise realization (Sec.~\ref{sm:rayleigh}). Here, we show how strongly the period doubling of the deterministic model depends on this phase.

Figure~\ref{fig:sm_phase} shows the correlation time $\tcorr$ of the deterministic response versus $\phi_0$, normalized to its value at $\phi_0=0$. The correlation time is largest near $\phi_0=0$ and vanishes near $\phi_0=\pi$, where the stroboscopic sampling times $t=n\tosc$ fall half a drive period out of step with the kick sequence. The drive phase therefore strongly affects the lifetime of the period doubling at both frequencies. This phase dependence, however, weakens as the drive frequency increases. We take $\phi_0=0$ as the reference phase in the deterministic model of Fig.~\ref{fig:fig2}(a), since it gives the longest correlation time. The fraction of phases that reach this lifetime grows with the drive frequency.

\section{Time scales of the toy model of time-solitons}
\label{sm:toymodel}

Here, we show several analytic results for the toy model of Eq.~\eqref{eq:DTLequation}. 

Deep inside the ordered regime, the stroboscopic response of Eq.~\eqref{eq:DTLequation} is a period-doubled signal. We denote by $M^z(n\tp)$ the stroboscopic magnetization at period $n$, which should be understood as the expectation value of the operator $\hat M^z$ for all-polarized initial states in the quantum case. At $p=0$, the normalized staggered magnetization $s(n)=(-1)^nM^z(n\tp)/M^z(0)$ of the main text therefore stays at its initial value $s(0)=1$, while for $0<p<1/2$ it flips sign with probability $p$ at each period. Without loss of generality, we write the normalized staggered magnetization $s(n)$ of a single realization as
\begin{equation}
s(n) = e^{-n\tp/\tth} \sigma(n), \quad \sigma(n)=\pm1 ,
\label{eq:sm_toy_s}
\end{equation}
where $\tp$ is the period of the drive, so that period $n$ occurs at the time $t=n\tp$. The factor $e^{-n\tp/\tth}$ is the prethermal envelope we introduced to capture the decay of the period doubling due to thermalization with a lifetime
\begin{equation}
  \tth \gg \tp . \label{eq:tth_limit}
\end{equation}
Here, $\sigma(n)$ is the subharmonic phase, which takes the two values $\pm1$ of the two period-doubled configurations. A time-soliton is the event $\varepsilon_n=\pi$ in the drive of Eq.~\eqref{eq:DTLequation}, which exchanges the sign of $\sigma(n)$, so that $\sigma$ follows the process
\begin{equation}
\sigma(n) = \sigma(n-1)\times
\begin{cases}
-1 & \text{prob. } p\\
+1 & \text{prob. } 1-p .
\end{cases}
\label{eq:sm_toy_sigma}
\end{equation}
The solitons are independent between periods. The gap $k$ between two neighboring solitons therefore has distribution $P(k)=(1-p)^{k-1}p$, with mean $\llbracket k\rrbracket=1/p$ periods, where $\llbracket\cdot\rrbracket$ denotes the average over the random realizations of the solitons. The typical time between two solitons is therefore $\llbracket k\rrbracket\tp$, and the density of time-solitons is its inverse,
\begin{equation}
\nsol \equiv \frac{1}{\llbracket k\rrbracket\tp} = \frac{p}{\tp} .
\label{eq:sm_nsol}
\end{equation}

Because the solitons are independent, the correlator of $\sigma(n)$ follows immediately. Over a time separation $k$, the sign is reversed once per soliton, so that $\sigma(m)\sigma(m+k)=(-1)^{N_k}$, with $N_k$ the number of solitons in the this interval. The averaged correlator thus is 
\begin{equation}
\llbracket \sigma(m)\sigma(m+k)\rrbracket = [(1-p)-p]^k = (1-2\nsol\tp)^k .
\label{eq:sm_sigma_corr}
\end{equation}

\subsection{Three time scales}

We can also define the three time scales of the main text in the context of the toy model. We consider 
\begin{align}
\tth &\equiv \tp\Bigl\llbracket \sum_{k\ge0} \bigl|s(k)\bigr| \Bigr\rrbracket ,
\label{eq:sm_tth_def}\\
\llbracket\tcorr\rrbracket &\equiv \tp\Bigl\llbracket \Bigl| \sum_{k\ge0} s(k) \Bigr| \Bigr\rrbracket ,
\label{eq:sm_tcorr1_def}\\
\tcorrens &\equiv \tp\Bigl| \sum_{k\ge0} \llbracket s(k)\rrbracket \Bigr| ,
\label{eq:sm_tcorrens_def}
\end{align}
where the prefactor $\tp$ converts the sum over periods into a time. Two of the three can be derived directly, as we show below. 

Using $|s(k)|=e^{-k\tp/\tth}$, Eq.~\eqref{eq:sm_tth_def} is nothing but
\begin{equation}
\tp\Bigl\llbracket \sum_{k\ge0}\bigl|s(k)\bigr|\Bigr\rrbracket = \frac{\tp}{1-e^{-\tp/\tth}}.
\label{eq:sm_tth_result}
\end{equation}
In the limit of Eq.~\eqref{eq:tth_limit}, it reduces to 
\begin{equation}
  \lim_{\tth\gg\tp} \frac{\tp}{1-e^{-\tp/\tth}} = \tth,
\end{equation}
which confirms that the integrated definition in Eq.~\eqref{eq:sm_tth_def} returns the prethermal scale $\tth$. 

Using the correlator of Eq.~\eqref{eq:sm_sigma_corr}, we can obtain
\begin{equation}
\tcorrens = \tp\sum_{k\ge0} e^{-k\tp/\tth}(1-2\nsol\tp)^k
= \frac{\tp}{1-(1-2\nsol\tp)e^{-\tp/\tth}}.
\label{eq:sm_tcorrens_result}
\end{equation}
In the limit of Eq.~\eqref{eq:tth_limit} and for $\nsol\tp\ll1$, it reduces to
\begin{equation}
  \lim_{\tth\gg\tp,\nsol\tp\ll1} \tcorrens
  = \frac{1}{2\nsol + \tth^{-1}}.
\end{equation}
Moreover, when the solitons dominate, $\nsol \gg \tth^{-1}$, this reduces to
\begin{equation}
\lim_{\tth^{-1} \ll \nsol \ll \tp^{-1}} \tcorrens = \frac{1}{2\nsol}.
\label{eq:sm_tcorr_2p}
\end{equation}
This is the limit where the soliton density alone sets the correlation time. 

The three time scales obey $\tcorrens\le\llbracket\tcorr\rrbracket\le\tth$, because the modulus taken inside a summation or ensemble average would give a result equal or larger than the modulus taken outside them. In the DTC regime, $\nsol\tth\ll1$, no soliton typically appears before thermalization. The modulus then act trivially, and three time scales collapse to one, 
\begin{equation}
  \tcorrens \simeq \llbracket\tcorr\rrbracket \simeq \tth ,
  \quad \nsol\tth\ll1 .
  \label{eq:sm_crystal_collapse}
\end{equation}
In the DTL regime, $\nsol\tth\gg1$, they separate, and $\tcorrens$ becomes directly relevant to the soliton density via Eq.~\eqref{eq:sm_tcorr_2p}.

\subsection{Distribution of the single-trajectory correlation}
\label{sm:toy_distribution}

The correlation time of single trajectory, before the modulus of Eq.~\eqref{eq:sm_tcorr1_def} is taken, is the random variable
\begin{equation}
  Z \equiv \tp\sum_{k\ge0} s(k) = \tp\sum_{k\ge0} a^k \sigma(k), \quad a \equiv e^{-\tp/\tth} ,
  \label{eq:sm_Zdef}
\end{equation}
a sum of the signs $\sigma(k)=\pm1$ weighted by the decaying envelope. Here, we derive the distribution for this random variable. 

Let $k_1<k_2<k_3<\dots$ be the period where the solitons appear in one trajectory. Between two neighboring solitons the sign does not change, so $\sigma(k)=+1$ for $k<k_1$, then $\sigma(k)=-1$ for $k_1\le k<k_2$, and so on, the sign alternating at each soliton. The sum in Eq.~\eqref{eq:sm_Zdef} can be written as 
\begin{equation}
  Z = \tp\left[\sum_{k=0}^{k_1-1}a^k - \sum_{k=k_1}^{k_2-1}a^k + \sum_{k=k_2}^{k_3-1}a^k - \dots \right].
  \label{eq:sm_Zsplit}
\end{equation}
Each block is a partial geometric series,
\begin{equation}
  \sum_{k=m}^{n-1}a^k = \frac{a^m-a^n}{1-a} ,
  \label{eq:sm_partialgeom}
\end{equation}
so that the block running from $k_i$ to $k_{i+1}$ contributes $(-1)^i\tp(a^{k_i}-a^{k_{i+1}})/(1-a)$. Summing them together, we obtain 
\begin{equation}
  Z = \frac{\tp}{1-a}\Bigl[1 - 2a^{k_1} + 2a^{k_2} - 2a^{k_3} + \dots\Bigr] .
  \label{eq:sm_Zblocks}
\end{equation}

It is convenient to separate that leading term by writing
\begin{equation}
  Z = \tp\,\frac{1+2W}{1-a} ,
  \qquad
  W \equiv \sum_{i\ge1}(-1)^i a^{k_i} ,
  \label{eq:sm_Wdef}
\end{equation}
so that all the randomness sits in $W$, and $W=0$ for a trajectory without solitons. 

\subsubsection{A self-consistent equation for the distribution of $W$}

By Eq.~\eqref{eq:sm_nsol}, each gap is an independent random variable $G$ with
\begin{equation}
  P(G=g) = (1-\nsol\tp)^{g-1}\nsol\tp , \qquad g=1,2,\dots
  \label{eq:sm_gap}
\end{equation}
Write the soliton positions as sums of the gaps, $k_i = g_1+g_2+\dots+g_i$ with each $g_i$ distributed as in Eq.~\eqref{eq:sm_gap}, and condition on the first gap, $k_1=g_1\equiv G$. We can express
\begin{align}
  W &= \sum_{i\ge1}(-1)^i a^{k_i} = \sum_{i\ge1}(-1)^i a^{k_1} a^{k_i-k_1} \nonumber\\
    &= -a^{G}\Bigl[1 - a^{k_2-k_1} + a^{k_3-k_1} - \dots\Bigr] .
  \label{eq:sm_Wfactored}
\end{align}
Consider now the bracket. Dropping its leading $1$, what remains is
\begin{equation}
  W' \equiv \sum_{j\ge1}(-1)^{j} a^{k_{j+1}-k_1} ,
  \label{eq:sm_Wprime}
\end{equation}
and Eq.~\eqref{eq:sm_Wfactored} reads 
\begin{equation}
  W = -a^{G}\bigl(1+W'\bigr) .
  \label{eq:sm_fixedpoint}
\end{equation}
This means that if the first soliton appears after $G=g$ periods (an event of probability Eq.~\eqref{eq:sm_gap}), then $W=-a^{g}(1+W')$. This allows us to write 
\begin{equation}
    P(W\le w) = \sum_{g\ge1}P(G=g) P(-a^{g}(1+W')\leq w).
    \label{eq:sm_cdf_conditioned}
\end{equation}

This random variable $W'$ has exactly the form of Eq.~\eqref{eq:sm_Wdef}, since we have $k_{j+1}-k_1 = g_2+\dots+g_{j+1}$, i.e.\ sums of the gaps $g_2,g_3,\dots$, in the same way that the positions entering $W$ are sums of $g_1,g_2,\dots$. To use this property, we focus on Eq.~\eqref{eq:sm_cdf_conditioned}, and rewrite the $-a^{g}(1+W')\leq w$ as $W' \ge -w a^{-g}-1$, and obtain 
\begin{align}
  P(W\le w) &= \sum_{g\ge1}P(G=g) P(W' \ge -w a^{-g}-1) \nonumber\\
  &= \sum_{g\ge1}P(G=g) \left[1-P(W' \le -w a^{-g}-1)\right] \nonumber\\
  &= \sum_{g\ge1}P(G=g) \left[1-P(W \le -w a^{-g}-1)\right]. 
  \label{eq:sm_cdf_equation}
\end{align}
This is a relation between $P(W\le w)$ and $P(W \le -w a^{-g}-1)$.

\subsubsection{Numerical solution}

We solve Eq.~\eqref{eq:sm_cdf_equation} numerically by iteration, using the following procedure.
\begin{enumerate}
  \item \emph{Grid.} Since $G\ge1$ and $W\le0$, Eq.~\eqref{eq:sm_fixedpoint} gives $a^G\in(0,a]$ and $1+W'\in[0,1)$, so the support is bounded,
  \begin{equation}
    -a \le W < 0 ,
    \qquad
    \tp\,\frac{1-2a}{1-a} \le Z < \frac{\tp}{1-a} \simeq \tth ,
    \label{eq:sm_support}
  \end{equation}
  and a uniform grid of $N_w$ points on $[-a,0]$ covers it with no truncation.
  \item \emph{Gap sum.} Truncate the sum over $g$ at $g_{\max}$ with $(1-\nsol\tp)^{g_{\max}}$ below machine precision.
  \item \emph{Iterate.} Starting from the uniform distribution on $[-a,0]$, evaluate the right-hand side of Eq.~\eqref{eq:sm_cdf_equation} and replace the old distribution by the result. The rescaled argument $-wa^{-g}-1$ generally falls between two grid points, so that $P(W\le\cdot)$ is obtained by linear interpolation. It may also fall outside the support of Eq.~\eqref{eq:sm_support}, where the distribution takes value $0$ when below the support or $1$ when above it. 
  \item \emph{Stop.} Monitor $\delta_n\equiv\max_w|P_{n+1}(W\le w)-P_n(W\le w)|$ and stop once it falls below the tolerance. The distribution of $Z$ then follows from the change of variable of Eq.~\eqref{eq:sm_Wdef}.
\end{enumerate}

\begin{figure}
\centering
\includegraphics{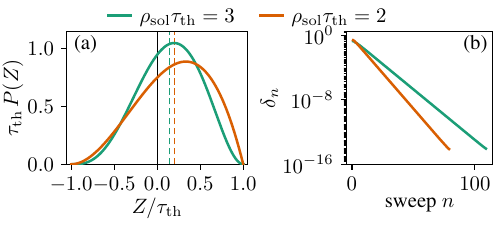}
\caption{
  \textbf{Distribution of $Z$ in the toy model.}
  (a) Distribution $P(Z)$ obtained from Eq.~\eqref{eq:sm_cdf_equation}, for two cases with $\nsol\tth$: $\tth=100$ with $\nsol=0.03$ ($\nsol\tth=3$), and $\tth=1000$ with $\nsol=0.002$ ($\nsol\tth=2$).  Dashed vertical lines mark the mean value from Eq.~\eqref{eq:sm_tcorrens_result}. A finite fraction of the weight lies at $Z<0$, coming from the trajectories with an odd number of solitons within the lifetime.
  (b) Convergence of the iteration, as the largest change per sweep $\delta_n$ against the sweep index $n$. 
  Both panels use a grid of $N_w=4001$ points and a tolerance of $10^{-14}$.
}
\label{fig:sm_toy_distribution}
\end{figure}

Figure~\ref{fig:sm_toy_distribution}(a) shows the distribution obtained in this way. Figure~\ref{fig:sm_toy_distribution}(b) shows that the iteration converges quickly. The distribution is broad, spanning the whole support, and carries appreciable weight on both sides of $Z=0$, the negative weight coming from the trajectories with an odd number of solitons within the lifetime. This is what separates the two correlation times of Eqs.~\eqref{eq:sm_tcorr1_def} and~\eqref{eq:sm_tcorrens_def}.

\begin{figure}
\centering
\includegraphics{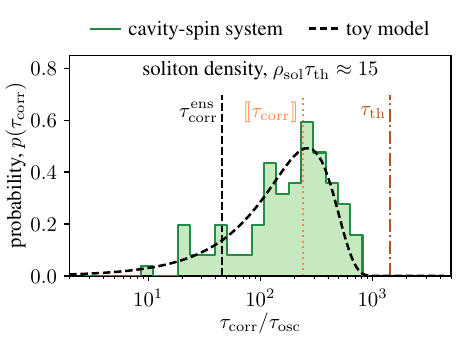}
\caption{
  \textbf{Distribution of the single-trajectory correlation time at a longer coherence time.}
  Same as Fig.~\ref{fig:fig4}, at $\tcav/\tosc=40$. Probability distribution $p(\tcorr)$, with $\tcorr=|Z|$, for the cavity-spin system (green histogram) and for the toy model with the parameters of that same lattice point (dashed line). The toy parameters are not fitted: $\tth$ is the value measured on the lattice, and $\nsol$ follows from inverting Eq.~\eqref{eq:sm_tcorrens_result} for the measured $\tcorrens$, which gives $\nsol\tth\approx15$ (top). The vertical lines mark $\tcorrens$, $\llbracket\tcorr\rrbracket$, and $\tth$. The toy model again reproduces the distribution. All other parameters are as in Fig.~\ref{fig:fig3}.
}
\label{fig:sm_toy_vs_lattice}
\end{figure}

The same construction can be tested against the spin lattice in cavity. For a given coherence time of the cavity mode, the lattice supplies both $\tth$ and $\tcorrens$ directly, and inverting Eq.~\eqref{eq:sm_tcorrens_result} for the latter fixes the soliton density $\nsol$. The toy model is then fully determined, and its distribution can be compared with the distribution of single-trajectory correlation times measured on the lattice. Figure~\ref{fig:fig4} of the main text shows the comparison deep in the DTL, at $\tcav/\tosc=20$ ($\nsol\tth=37$). Figure~\ref{fig:sm_toy_vs_lattice} shows it at $\tcav/\tosc=40$ ($\nsol\tth=15$). Both figures show the single-trajectory correlation time $\tcorr=|Z|$ on a logarithmic axis. In both cases the distribution is broad, spanning more than an order of magnitude in $\tcorr$, and is peaked well below the prethermal scale. The toy model captures the position of the peak and the overall width, and reproduces the slow rise on the small-$\tcorr$ side.

\section{Phase diagram}
\label{sm:classification}

The phase diagram of Fig.~\ref{fig:fig1}(c) is classified by three quantities, defined in the main text and in Methods. We recall them here in terms of the normalized staggered magnetization $s_r(n)=(-1)^nM^z_r(n\tosc)/M^z_r(0)$ of realization $r=1,\dots,N_{\rm traj}$, which satisfies $s_r(0)=1$. The three quantities are the noise-averaged single-trajectory correlation time, the thermalization time, and the precession weight,
\begin{align}
\llbracket\tcorr\rrbracket &= \frac{\tosc}{N_{\rm traj}}\sum_{r=1}^{N_{\rm traj}}\Bigl|\sum_{n=0}^{N_p} s_r(n)\Bigr| ,
\label{eq:sm_class_tcorr}\\
\tth &= \frac{\tosc}{N_{\rm traj}}\sum_{r=1}^{N_{\rm traj}}\sum_{n=0}^{N_p} \bigl|s_r(n)\bigr| ,
\label{eq:sm_class_tth}\\
P_{\rm prec} &= \max\bigl(0,-\min_k \mathcal C(k)\bigr),
\label{eq:sm_class_Pprec}
\end{align}
where $\mathcal C(k)=\frac{\sum_{m}\llbracket s(m+k) s(m)\rrbracket}{\sum_{m}\llbracket s(m)^2\rrbracket}$, and $m$ runs over a reference window of $100$ periods starting at $m=50$. Each parameter point is then labeled in three steps. Points in the freely precessing band at low temperature, where $P_{\rm prec}$ has not yet fallen to its high-temperature floor $P_{\rm floor}$, show free precession. Among the remaining points, those with $\llbracket\tcorr\rrbracket<10\tosc$ are disordered. All other points are a DTC if $\llbracket\tcorr\rrbracket/\tth>0.418$ and a DTL otherwise. Figure~\ref{fig:sm_orderparams} shows the three quantities over the plane of Fig.~\ref{fig:fig1}(c). The levels $P_{\rm floor}$ and $0.418$ are fixed by the data, as described below, whereas the level $10\tosc$ is a choice.

\begin{figure*}
\centering
\includegraphics{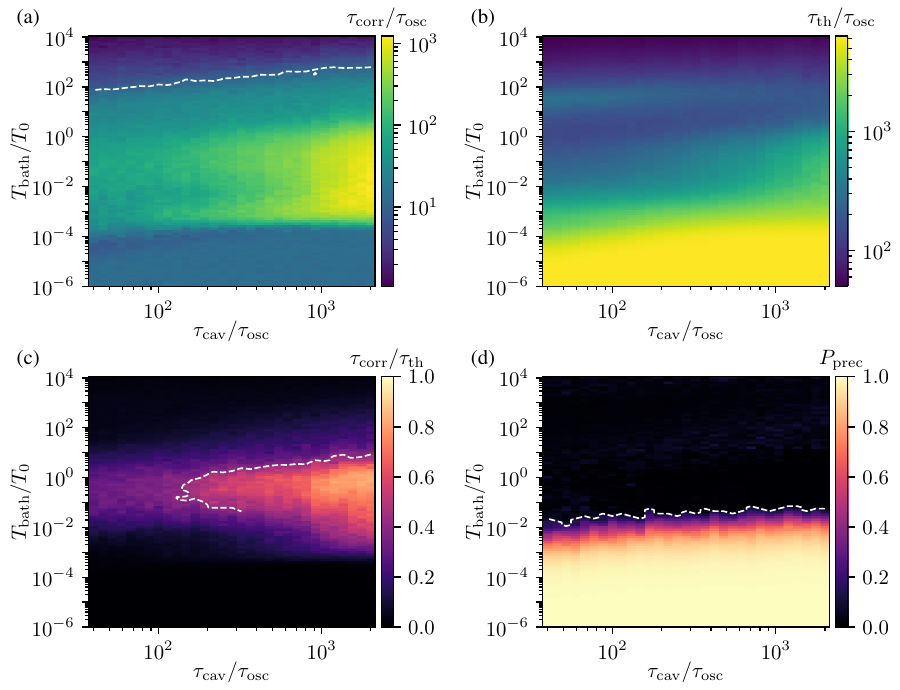}
\caption{
  \textbf{The quantities behind the classification, over the plane of Fig.~\ref{fig:fig1}(c).}
  (a) Correlation time $\llbracket\tcorr\rrbracket$ and (b) thermalization time $\tth$. (c) Their ratio and (d) the precession weight $P_{\rm prec}$, on linear scales. Each panel carries, as a white dashed line, the boundary that its own quantity defines: the disorder level $\llbracket\tcorr\rrbracket=10\tosc$ in (a), the DTC level $\llbracket\tcorr\rrbracket/\tth=0.418$ in (c), and the edge of the freely precessing band in (d). Panel (b) defines no boundary. Comparing (a) and (b) shows the origin of the DTC island. The thermalization time $\tth$ falls smoothly with temperature everywhere, whereas $\llbracket\tcorr\rrbracket$ rises to meet it only within a window that widens with the coherence time of the cavity mode. Parameters as in Fig.~\ref{fig:fig1}(c). See text for more details.
}
\label{fig:sm_orderparams}
\end{figure*}

We classify with the single-trajectory scale $\llbracket\tcorr\rrbracket$ rather than with the ensemble scale $\tcorrens$, since the DTC and the DTL are distinguished by how long a single realization keeps its subharmonic phase.

The precession weight $P_{\rm prec}$ stays at unity as long as the drive is too weak to disturb the free precession of the spins. 
It then falls as the temperature increases and settles on a floor of $0.02$ to $0.03$, set by the residual noise of the correlator. 
We determine $P_{\rm floor}$ from this floor in each column of fixed $\tcav$. 
Specifically, we collect the values of $P_{\rm prec}$ at the highest temperatures, from one tenth of the maximum temperature upward.
We set $P_{\rm floor}$ to their median plus five times their median absolute deviation.
The free precession then extends from the lowest temperature up to the temperature at which $P_{\rm prec}$ first reaches $P_{\rm floor}$.

To fix the cut between the DTC and the DTL, we consider all points that show local temporal order, i.e., all points that are neither freely precessing nor disordered.
The histogram of $\llbracket\tcorr\rrbracket/\tth$ over these points has two groups.
Most points have a small ratio, around $0.12$, and form the DTL.
A smaller group has a ratio up to $0.83$, where $\llbracket\tcorr\rrbracket$ is comparable to $\tth$, and forms the DTC.
The two groups are separated by a dip near $0.4$.
We place the cut at the value that best separates the two groups, $0.418$.
Because the dip is shallow, the number of DTC points depends on the exact cut.

The boundary between the DTL and the disordered region is the least sharp of the three. The distribution of $\llbracket\tcorr\rrbracket$ shows no feature near the cut $\llbracket\tcorr\rrbracket=10\tosc$. Moving the cut between $3$ and $30$ cavity periods changes the number of disordered points from $256$ to $816$. This is expected, since the local temporal order is lost gradually on heating.

\section{Spin backaction on the oscillator}\label{sm:backaction}

Here, we demonstrate the backaction effect by including the backaction term of Sec.~\ref{sm:thermo}. The intensive oscillator equation then reads
\begin{equation}
\ddot y + \gamma\dot y + \omega_0^2 y
= \xi(t) - \frac{\lambda}{M_{\rm osc}}\,h^-(t).
\end{equation}
The per-spin force $h^-(t)=H^{(-)}(t)/N$ is defined in Sec.~\ref{sm:thermo}, with $H^{(-)}$ given by Eq.~\eqref{eq:sm_Hminus}. 

\begin{figure*}
\centering
\includegraphics{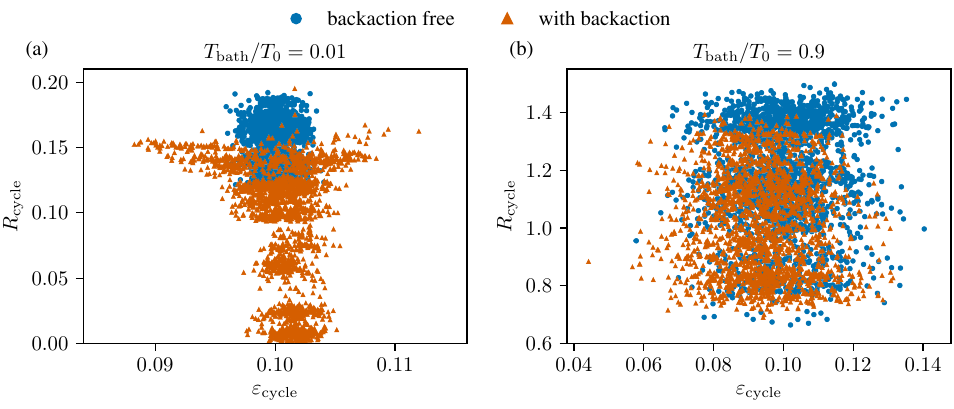}
\caption{
  \textbf{Effect of spin backaction.} Per-cycle effective-drive parameters $(\varepsilon_{\rm cycle},R_{\rm cycle})$ for single trajectories at (a)~$\Tbath/T_0=0.01$ and (b)~$\Tbath/T_0=0.9$. Blue circles show the backaction-free trajectories used in Fig.~\ref{fig:fig2}(h,i), and orange triangles show the corresponding trajectories with backaction. At each temperature, the two trajectories use the same noise sequence and initial spin configuration. All other parameters are those of Fig.~\ref{fig:fig2}.
}
\label{fig:sm_backaction}
\end{figure*}

We repeat the calculation of Fig.~\ref{fig:fig2}(h,i) with the backaction term included and compare the per-cycle effective-drive parameters $(\varepsilon_{\rm cycle},R_{\rm cycle})$ in Fig.~\ref{fig:sm_backaction}. At both $\Tbath/T_0=0.01$ and $\Tbath/T_0=0.9$, the backaction reshapes the cloud, shifting its weight toward smaller $R_{\rm cycle}$ and broadening its distribution along $\varepsilon_{\rm cycle}$. These results indicate that the spins can reshape the noisy drive. How this feedback modifies the phase diagram of Fig.~\ref{fig:fig1}(c) remains an open question.

\section{Numerical methods}\label{sm:numerics}

\begin{figure}
\centering
\includegraphics{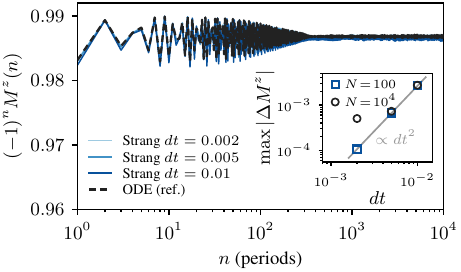}
\caption{
  \textbf{Integrator convergence.} Deterministic drive, $\tosc=0.2$. Main panel: the stroboscopic subharmonic response $(-1)^n M^z(n\tosc)$ from the exact-rotation Strang splitting at $dt=0.002,0.005,0.01$ and from the adaptive ODE solver (black dashed, reference), which coincide over $10^4$ periods. Inset: the Strang error $\max|\Delta M^z|$ relative to the ODE reference versus $dt$, over the first $100$ periods (squares) and over the full $10^4$-period run (circles). The early error follows the second-order $\propto dt^2$ scaling (gray line). The full-run error stays bounded and saturates near the reference's own accuracy floor ($\sim5\times10^{-4}$), confirming the integrator over the whole run.
}
\label{fig:sm_numerics}
\end{figure}

We evolve the classical spins by the Poisson-bracket equation $\dot{\vec S}_i=\vec S_i\times\vec B_i$, with the instantaneous local field $\vec B_i=\bigl(g_{\rm eff}(t),\,0,\,J_{\rm eff}(t)\sum_{j\in{\rm nn}(i)}S^z_j\bigr)$. It is integrated by a second-order Strang splitting into a transverse-field sector (every spin rotates about $\hat x$ by $g_{\rm eff}\,dt$) and an Ising sector (each spin rotates about $\hat z$ by $J_{\rm eff}\sum_j S^z_j\,dt$, with all $S^z_i$, and hence every $z$-field, frozen during the substep). Each substep is an \emph{exact} rotation, so $|\vec S_i|=1$ is preserved to machine precision and the scheme is stable over the $10^4$ periods used here. The oscillator coordinate $y(t)$ is propagated on the same time step and held fixed within it. Each spin is initialized in a spherical cap of half-angle $\theta_{\rm init}$ about $+\hat z$ (see Methods), from a fixed random seed.

We validate the splitting against an independent high-order adaptive ODE solver. Figure~\ref{fig:sm_numerics} shows that the subharmonic response is indistinguishable across Strang steps $dt\in\{0.002,0.005,0.01\}$ and the ODE reference over $10^4$ periods, and that the error versus the reference follows the expected second-order $\propto dt^2$ scaling at early times (inset). Over the full run, the error remains bounded at $\sim5\times10^{-4}$ for the smaller steps. This value is the accuracy of the reference solver itself, whose solutions at successive tolerances already differ by this much. No residual splitting error is therefore resolvable below this level, and the production step $dt=0.005$ lies well within the converged regime.

\fi 


%

\end{document}